\documentclass[
twocolumn,
superscriptaddress,
 amsmath,amssymb,
 aps,
prr,
floatfix,
longbibliography
]{revtex4-2}

\usepackage[utf8]{inputenc}
\usepackage[T1]{fontenc}

\usepackage{CJK}
\usepackage{multirow}
\usepackage{tabularx}
\usepackage{graphicx}
\DeclareGraphicsExtensions{.jpg,.png}
\usepackage{capt-of}
\usepackage[colorlinks=true, citecolor=blue, linkcolor=blue]{hyperref}

\begin{document}

%\title{The role of inter-particle cohesive stiffness in determining the size and nature of self-limiting, geometrically-frustrated assemblies}

\title{Shape equilibria of vesicles with rigid planar inclusions}

\author{Geunwoong Jeon}
\affiliation{Department of Physics, University of Massachusetts, Amherst, Massachusetts 01003, USA}

\author{Justin Fagnoni}
\affiliation{Department of Physics, University of Massachusetts, Amherst, Massachusetts 01003, USA}

\author{Hao Wan}
\affiliation{Department of Polymer Science and Engineering, University of Massachusetts, Amherst, Massachusetts 01003, USA}

\author{Maria M. Santore}
\affiliation{Department of Polymer Science and Engineering, University of Massachusetts, Amherst, Massachusetts 01003, USA}

\author{Gregory M. Grason}
\affiliation{Department of Polymer Science and Engineering, University of Massachusetts, Amherst, Massachusetts 01003, USA}

\date{\today}

\begin{abstract} % abstract
Motivated by recent studies of two-phase lipid vesicles possessing 2D solid domains integrated within a fluid bilayer phase, we study the shape equilibria of closed vesicles possessing a single planar, circular inclusion.  While 2D solid elasticity tends to expel Gaussian curvature, topology requires closed vesicles to maintain an average, non-zero Gaussian curvature leading to an elementary mechanism of shape frustration that increases with inclusion size.  We study elastic ground states of the Helfrich model of the planar-fluid composite vesicles, analytically and computationally, as a function of planar fraction and reduced volume.  Notably, we show that incorporation of a planar inclusion of only a few percent dramatically shifts the ground state shapes of vesicles from predominantly {\it prolate} to {\it oblate}, and moreover, shifts the optimal surface to volume ratio far from spherical shapes.  We show that for sufficiently small planar inclusions, the elastic ground states break symmetry via a complex variety of asymmetric oblate, prolate, and triaxial shapes, while inclusion sizes above about $8\%$ drive composite vesicles to adopt axisymmetric oblate shapes.  These predictions cast useful light on the emergent shape and mechanical responses of fluid-solid composite vesicles.  
\end{abstract}

\maketitle

\section{Introduction}\label{introsec}

Closed fluid membranes, such as lipid bilayer vesicles, constitute a paradigmatic class of shape-adaptive soft matter, with known studied connections to biological materials~\cite{dimova2019giant}.  Even single component vesicles are known to adopt a complex spectrum of low-symmetry shapes, which can be effectively modeled as equilibria of a continuum theory controlled by relatively few thermodynamic parameters, like surface-to-volume ratio and asymmetry between the inner and outer membrane\cite{seifert1997configurations}. Even for the simplest case of homogeneous vesicles without spontaneous curvature, the shape equilibria of closed free vesicles vary considerably with symmetry decreasing with progressive deflation, from spherical, to prolate, to oblate, and ultimately to stomatocyte shapes~\cite{seifert1991shape, ziherl2005nonaxisymmetric}. Recent work shows that these distinct equilibria exhibit vastly different responses to shape perturbation upon localized forcing \cite{reboucas2024stationary}.

\begin{figure}
\begin{center}
\includegraphics[width=\columnwidth]{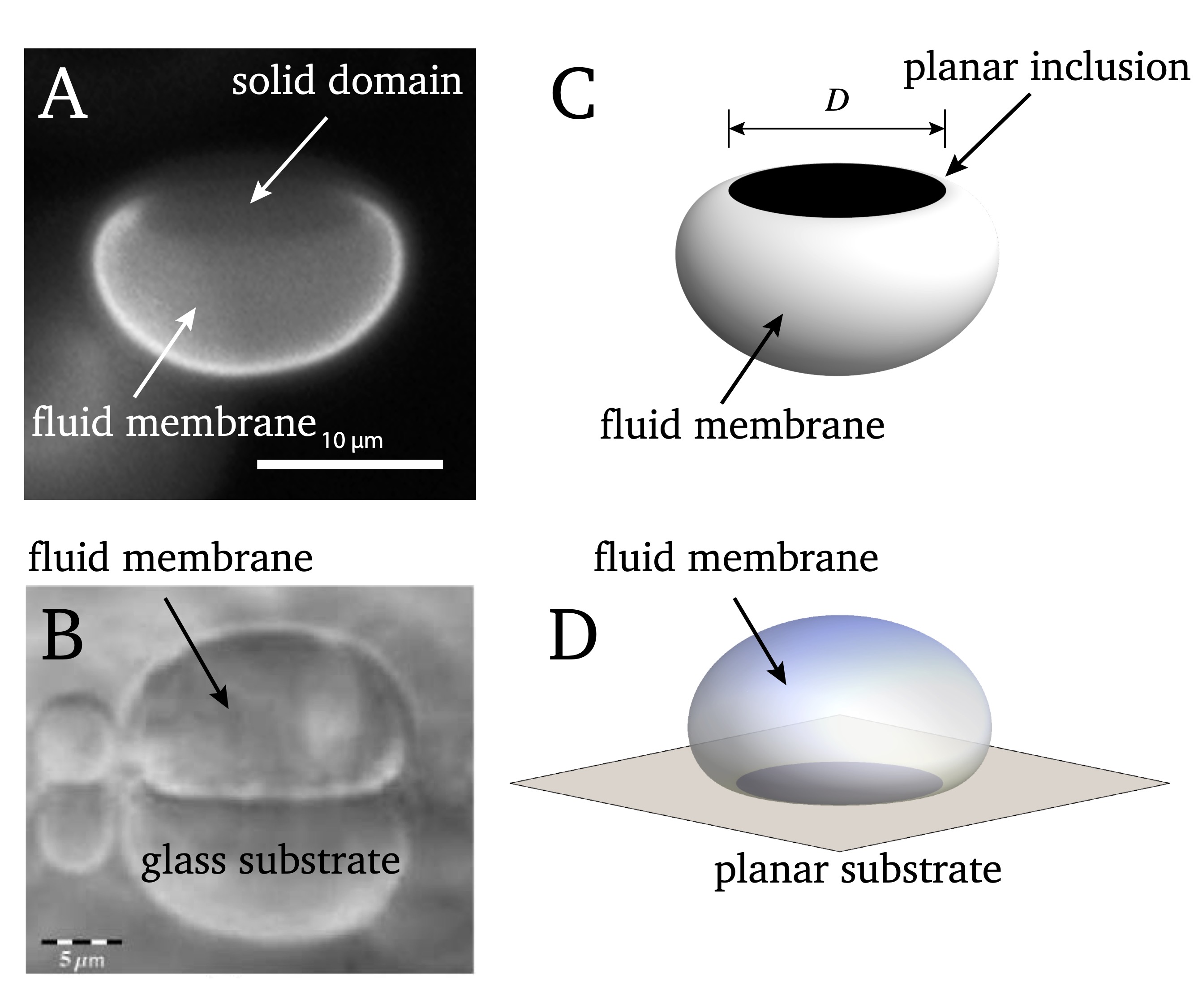}%
\end{center}
\caption{\label{fig: exam}(A) A microscopic image of a solid-fluid composite lipid vesicle (B) a microscopic image of a fluid vesicle adhered to glass substrate (Reprinted with permission from Gruhn et al., Langmuir, 2007\cite{gruhn2007novel}. Copyright 2007 American Chemical Society) (C) computational model of the fluid membrane with a circular inclusion (D) computational model of fluid membrane adhered to the planar substrate.}
\end{figure}

The possibilities for symmetry breaking are far more complex for inhomogeneous vesicles~\cite{lipowsky2002domains}. This includes fluid-fluid phase-separated vesicles, where mechanical properties (e.g. bending moduli, spontaneous curvature) vary with local composition~\cite{julicher1996shape, Baumgart_2003, Ursell_2009, Gutlederer_2009, lipowsky2024multispherical}, as well as closed membranes with protein or particulate absorbates and inclusions~\cite{Weikl_1998, Schweitzer_2015, Koltover_1999, vanderWel_2016}, which locally perturb membrane shape and may collectively reconfigure the global ground symmetries.  Among inhomogeneous vesicle categories, the shape equilibria of fluid-solid composites remain particularly poorly understood.  Recent studies of fluid-solid composite derive from multi-component lipid vesicles, where one component phase separates into a solid phase upon cooling from a mixed phase, consistent with one or more 2D solids domains coexisting within an otherwise homogeneous 2D ``background'' fluid.  Observations find that solid domains may vary in number, size, and shape depending on both equilibrium and non-equilibrium processing conditions~\cite{chen2014large, lipowsky2002domains, Bandekar_2012}.  

Generic geometric and mechanical considerations complicate the understanding of fluid-solid composite vesicle shape.  On one hand,  membrane bending elasticity alone tends to favor uniform mean-curvature shapes, which for closed vesicles is optimal for spherical shapes.   On the other hand, 2D solid (plate) elasticity strongly resists spherical curvature, due to the coupling between Gaussian curvature and in-plane strains~\cite{Witten_2007, Grason_2016}.  While the mean value of Gaussian curvature of closed vesicles is fixed and positive, simple considerations of the relative energetics of solid strain energy in comparison to membrane bending energy suggest that Gaussian curvature is strongly partitioned in fluid-solid composite vesicles.  The in-plane strain energy needed to force a solid domain of size $D$ onto a sphere of size $R$ is proportional to $Y ~ D^6/R^4$, where $Y$ is the 2D Youngs modulus\cite{Schneider_2005, Davidovitch_2019}.  In comparison, the characteristic bending energy cost is proportional to $B ~ D^2/R^2$, where $B$ is the bending modulus~\cite{helfrich1973elastic}.  Notably, the ratio $\sqrt{B/Y}\equiv t $ defines an elastic thickness that is comparable in dimensions to the nanometer scale thickness of a 2D solid, while $R$ is of the order 10s of microns for GUVs (Giant Unilamellar Vesicles)~\cite{dimova2019giant}. Hence, for a solid domain size comparable to the vesicle size, the ratio of strain to bending energy grows asymptotically as large with vesicle size as $(Y/B) D^4/R^2 \sim (R/t)^2 \sim 10^8$.  This scaling implies that forcing the solids to adopt the spherical curvature imposed by the closed vesicle is prohibitive relative to the cost of redistributing that Gaussian curvature to the fluid phase, where it incurs only additional bending energy at the scale of $\sim B$.  On these grounds, it can be expected that solid domains on composite vesicles adopt {\it developable shapes}~\cite{Witten_2007}, with the surrounding fluid phase taking up the appropriate increase of Gaussian curvature required by the fixed topology. Hence, we may view the expulsion of Gaussian curvature from the solid domains into the surrounding fluid phase and resulting inhomogeneous curvatures as a mechanism of {\it shape frustration}, and the magnitude of that frustration will tend to increase with increasing solid domain fraction $\Phi$.

\begin{figure}[t!]
\begin{center}
\includegraphics[width=0.95\columnwidth]{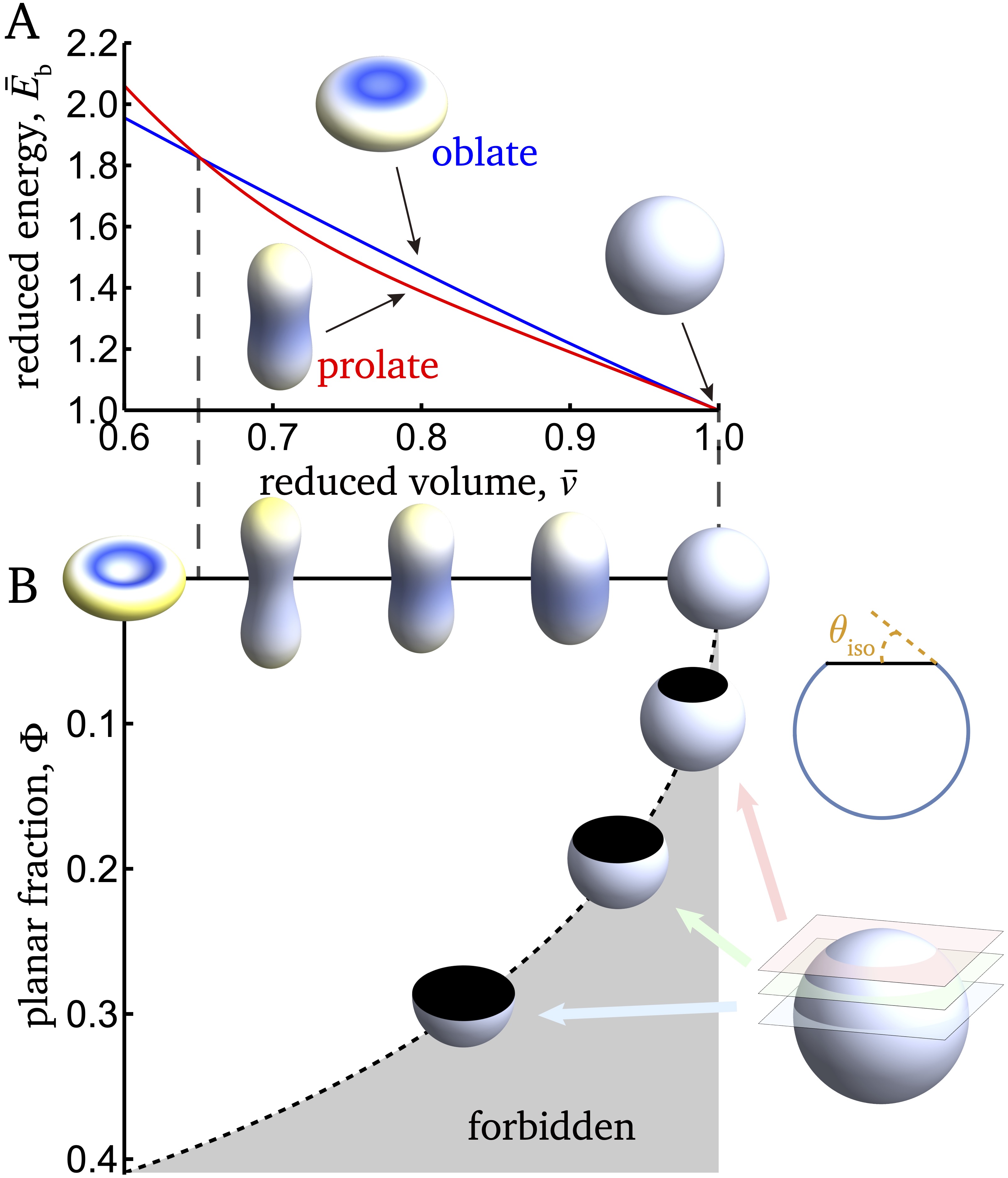}%
\end{center}
\caption{\label{fig: param}(A) Competing prolate and oblate branches for ground state shapes of homogeneous vesicles as a function of reduced volume.  (B) Schematic representation parameter space for single-planar inclusion, where the maximal inflation, isoperimetric limit, separates allowed from forbidden regimes of shape space illustrated by examples on the right side.}
\end{figure}

In this article, we study a minimal model of this shape frustration, where we consider a fluid membrane coexisting with a single rigid, planar, and circular inclusion in a closed, composite vesicle. While solid domains may in fact bend at low energy with Gaussian curvature (i.e. isometrically), the case of a planar inclusion presents a tractable setting for studying structure and thermodynamics over a comprehensive range of inflations and planar fractions.  Moreover, as we show, the gross features of shape equilibria for the rigid planar model agree well with experimentally observed shapes of fluid-solid composites with large, compact solid domains (see example in Fig.~\ref{fig: exam}A,C).  Additionally, we note that the problem of a fluid vesicle in mechanical equilibrium with a planar region is also fundamental to models of vesicle adhesion on planar substrates (see Fig.~\ref{fig: exam}B,D), where binding transition and resultant bending energy of the bound vesicle both depend on the (fixed) metric of the substrate\cite{seifert1991adhesion,lipowsky1991adhesion,das2008adhesion,li2023curvature}.

Before proceeding to our model of composite vesicles with rigid, planar inclusions, we briefly review the classical results of shape equilibria for homogeneous vesicles, which constitutes the limiting $\Phi \to 0$ case of our planar inclusion study.  The axisymmetric ground states of the Helfrich model for homogeneous vesicles were first studied by the analytical solution of the shape equations~\cite{seifert1991shape, seifert1997configurations}, and subsequent numerical studies~\cite{jie1998numerical, ziherl2005nonaxisymmetric, bian2020bending} show that ground state shapes retain axisymmetry in the absence of spontaneous curvature.  Two solution branches compete in the regime of high inflation, measured by the {\it reduced volume} parameter
\begin{equation}
    \bar v\equiv\sqrt{36\pi}\frac{V}{A^{3/2}}
\end{equation}
where $V$ and $A$ are respective volume and area, shown in Fig.~\ref{fig: param}A, the reduced energy ratio relative to a perfectly spherical shape.  The prolate branch is lowest energy for $\bar{v} \geq 0.65$, while oblate shapes have lower energy at lower inflation\footnote{For homogeneous vesicles without spontaneous curvature, stomatocyte shapes are lower energy than both prolate and oblate branches for especially low inflation $\bar{v}<0.59$.}.  Both branches continuously meet the minimal energy spherical shape at maximal inflation as $\bar{v} \to 1$.  In what follows we consider the symmetry (e.g. oblate vs. prolate) of ground state shapes in an expanded parameter space which includes the {\it planar fraction}
\begin{equation}
    \Phi\equiv\frac{A_\text{planar}}{A}
\end{equation}
where $A_\text{planar} = \pi D^2/4$ is the area of the planar disc that continuously meets the fluid vesicle portion.  Before consideration of mechanical equilibria, we note that a finite planar inclusion reduces the range of accessible inflation.  It is straightforward to understand the states of maximum inflation, known as the isoperimetric limit shapes, because the maximal volume that can be enclosed by the fluid portion remains  spherical.  Hence, the isoperimetric limit shapes are simply spheres cut by planes (see Fig.~\ref{fig: param}B), whose intersections enclose the appropriate disc area.  It is straightforward to determine the relationship between planar fraction and the maximal reduced volume
\begin{equation}\label{isoperieq}
    \bar v_\text{max}(\Phi)=\sqrt{1-2\Phi}(1+\Phi),
\end{equation}
which notably is below the spherical limit for any finite $\Phi$ and vanishes completely at $\Phi =1/2$, imposing any upper bound for the planar fraction in our study, shown as the dashed line in Fig.~\ref{fig: param}B.  The isoperimetric limit shapes are not smooth, since the fluid and inclusion meet at a finite angle $\theta_{\rm iso}$.  We show below that the diverging curvature in this limit strongly influences the symmetry and energy landscape of ground state shapes.

In this article, we investigate ground state shapes of the planar inclusion model in the parameter regime highlighted in Fig.~\ref{fig: param}B via a combination of analytical and finite-element solutions to shape equilibrium.  In Sec.~\ref{sec: methods} we summarize our approach to the axisymmetric solutions of the shape equations for the planar inclusion model, as well as our approach to non-axisymmetric shapes via SE (Surface Evolver).  We first discuss the evolution of prolate and oblate solution branches for the axisymmetric case in Sec.~\ref{sec: axisymmetry} with increasing planar fraction.  We show that increasing $\Phi$ shifts the relative stability of these two branches, with shapes in the prolate branch completely disappearing from the axisymmetric ground states for $\Phi > 0.023$.  We also show that minimal energy shapes fall below the upper isoperimetric limit of inflation, and hence, the incorporation of a planar inclusion leads to regimes of positive and negative membrane tension.  We next consider the ground state shapes without any assumption of axisymmetry via the Surface Evolver model in Sec.~\ref{sec: nonaxi}.  Indeed, we show that for sufficiently low planar fraction, ground states break axisymmetry and exhibit a range of complex quasi-prolate and quasi-oblate shapes, while for sufficiently large planar fraction ($\Phi \gtrsim 0.08$), and notably near to zero tension, ground states retain axisymmetric, oblate shapes.  Next, in Sec.~\ref{sec: nearlyiso}, we analyze the asymptotic approach to the maximal inflation shapes and argue that morphological frustration imposed by a planar inclusion leads to a characteristic concentration of diverging fluid bending energy at the boundary between the phases. We conclude in Sec.~\ref{sec: conclusion} with a discussion of the relevance of the planar inclusion model for experimental shapes of fluid-solid composite vesicles, as well as remaining questions about the role of the flexibility of solid domains on the ground state shape phase diagram of this system.

\section{Model of fluid vesicle with a planar inclusion}
\label{sec: methods}

Here we introduce our model of composite fluid-planar inclusion vesicles.  We consider closed vesicles including a single, rigid disc region (i.e. inclusion) which smoothly matches the tangents of coexisting fluid.  Tangent continuity is required for finite bending energy at the boundary of fluid-solid membranes, as well as at the edge of a planar contact zone of an adhering vesicle.  For the case of rigid inclusions, the elastic energy derives entirely from the fluid bending energy which has the form of Helfrich energy\cite{helfrich1973elastic},
\begin{equation}
    E_\text{b}=\frac{B}{2}\oint_{\rm fluid} {\rm d} A\ (2H)^2+B_{\rm G}\oint{\rm d} A\ K_{\rm G}
    \label{eq: helf}
\end{equation}
where $H$ and $K_{\rm G}$ are the respective mean and Gaussian curvature of the fluid, whose elasticity is parameterized by the respective bending moduli, $B$ and $B_{\rm G}$.  Because the fluid membrane is taken to meet continuously tangent to the planar inclusion and the composite membrane is closed, the second term in eq.~(\ref{eq: helf}) is topologically invariant and equal to a constant $4 \pi B_{\rm G}$ for all shapes~\cite{Kamien_2002}.  Hence, we neglect this term in the remaining analysis.  Additionally, we consider the ensemble of fixed areas of fluid and planar inclusions, as well as fixed internal volume of the membrane.

\begin{figure}
\begin{center}
\includegraphics[width=0.8\columnwidth]{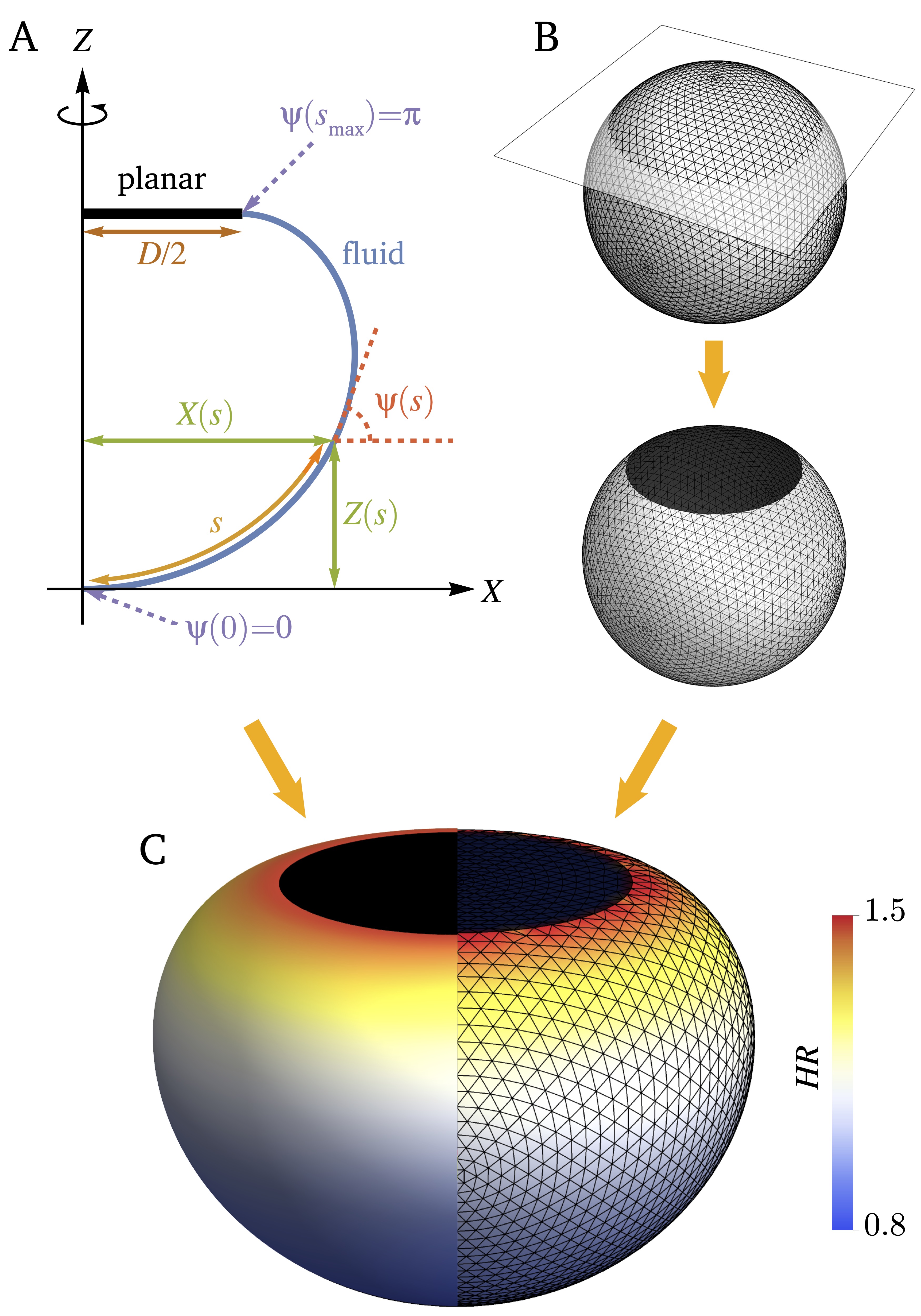}%
\end{center}
\caption{\label{fig: methods}The arclength parametrization of the membrane shape under global axisymmetry. $s$ is the arclength along the membrane, $X(s)$ and $Z(s)$ are the spatial coordinates, and $\psi(s)$ is the tangent angle.}
\end{figure}

Fig.~\ref{fig: methods} shows schematically the two methods we employed to study equilibria of this model, as detailed below:  A) solutions of axisymmetric shape equations and B) Surface Evolver simulations of vesicles with fixed planar regions.

\subsection{Shape equations and axisymmetric parameterization}

\label{sec: aximethod}

Here, we study the equilibria of the fluid bending energy under the constraints of fixed inclusion size, fixed fluid phase area $A_{\rm fluid}$ and fixed enclosed volume $V$.  Constraining the latter two quantities is accomplished by including Lagrange multipliers coupled to $A_{\rm fluid}$ and $V$ (i.e. membrane tension and pressure) in an augmented thermodynamic potential
\begin{equation}
    F=E_\text{b}+\Sigma A_{\rm fluid}-PV\ \label{freeenergy}
\end{equation}
which is a function of the vesicle shape.  The equilibrium shapes can be found by minimizing the membrane energy with respect to variation of the fluid membrane shape (i.e. $\delta F=0$).  This gives the standard shape equation of the Helfrich model (in the absence of spontaneous curvature)\cite{zhong1989bending,seifert1997configurations}
\begin{equation}
    -P+2\Sigma H-2B\big[2H(H^2-K_{\rm G})-\nabla_\perp^2 H \big]=0\label{eq: shapeeq} ,
\end{equation}
where $\nabla_\perp^2$ is the surface Laplacian operator.  Following ref.~\cite{seifert1991shape}, as shown in Fig.~\ref{fig: methods}A, we implement and solve this equation in terms of the curve $\{X(s), Z(s) \}$ in the plane parallel to the rotation axis, and $s$ is an arclength parameter running from $s=0$ on the bottom pole to $s=s_{\rm max}$ at the edge of the solid domain.  As shown in Fig.~\ref{fig: methods}A $Z(s)$ is the position along the rotation axis and $X(s)$ is the radial separation of the fluid membrane at $s$ from the symmetry axis.  The arc length derivatives of position in this plane define the orientation angle $\psi(s)$
\begin{equation}
    X'(s) = \cos \psi(s); \  Z'(s) = \sin \psi(s) .
\end{equation}
Based on these coordinates, the shape equation becomes a set of coupled ODEs, as detailed in Appendix~\ref{app: exp}.  These equations are solved subject to the boundary conditions of continuity at the fluid pole of the vesicle
\begin{equation}
    \label{eq: bot}
    X(0) = 0; \ \psi(0) = 0 ,
\end{equation}
and at the edge of the planar inclusion
\begin{equation}
    \label{eq: top}
    X(s_{\rm max}) = D/2; \ \psi(s_{\rm max}) = \pi ,
\end{equation}
where again we assume tangent continuity, and hence, finite curvature at the planar-fluid boundary.  We implement a shooting-method approach in Mathematica to solve the shape equations, eq.~\ref{axishapeeq}, boundary conditions, and constraints for fixed fluid/planar area and enclosed volume, in terms of unknown parameters (e.g. $\Sigma$, $P$ and $s_{\rm max}$).

Resulting equilibria are analyzed below in terms of shape (curvature) and energetics (see example on left half of Fig.~\ref{fig: methods}C).  For the purposes of analysis, it is helpful to work in terms of rescaled variables.  In particular, we consider energetic quantities normalized by $B$,
\begin{equation}
\bar{E}_{\rm b} \equiv \frac{E_{\rm b}}{8 \pi B}; \ \bar{\Sigma} \equiv \Sigma/B; \ \bar{P} \equiv P/B.
\end{equation}
Note that reduced energy, $\bar{E}_{\rm b}$, is defined such that the energy of a homogeneous sphere is 1.  We also define the characteristic radius $R$ in terms of an equal volume sphere, to characterize to macroscopic size of the shape solutions,
\begin{equation}
R\equiv \sqrt{\frac{A}{4 \pi}}.
\end{equation}
Notably, as with homogeneous vesicles in the absence of spontaneous curvature, the energy of the fluid-planar inclusion composite model is scale invariant.  Rescaling the macroscopic size $R$ while holding $\Phi$ and $\bar{v}$ fixed does not change $\bar{E}_{\rm b}$.  Using this fact and standard variational principles for vesicles with fixed area and volume~\cite{seifert1997configurations} yields the following relations 
\begin{equation}
 A \bar{\Sigma} = -\frac{A}{B} \Big( \frac{\partial E_{\rm b}}{ \partial A} \Big)_{V,\Phi} = 12 \pi \bar{v} \Big( \frac{\partial E_{\rm b}}{ \partial \bar{v}} \Big)_{\Phi} , 
  \label{eq: ten}
\end{equation}
and
\begin{equation}
 V \bar{P} = \frac{V}{B} \Big( \frac{\partial E_{\rm b}}{ \partial V} \Big)_{A,\Phi} = 8 \pi \bar{v} \Big( \frac{\partial E_{\rm b}}{ \partial \bar{v}} \Big)_{\Phi} ,
 \label{eq: press}
\end{equation}
which relate both internal tension and pressure to the same first derivative of reduced energy.

\subsection{Surface Evolver composite vesicle model}

\label{sec: SE}

% \begin{figure}[t!]
% \begin{center}
% \includegraphics[width=\columnwidth]{SEprotocol.jpg}%
% \end{center}
% \caption{\label{SEprotocol}The schematics of SE protocol that was introduced in the main text. The white faces are the fluid membrane and the navy faces are the inclusion.}
% \end{figure}
To complement axisymmetry solutions to the shape equations, we consider non-axisymmetric shapes via numerical minimization of a discrete-mesh implementation of composite vesicle energy in Surface Evolver\cite{brakke1992surface,brakke2013surface}. In this discrete description, the bending energy was computed based on the vertices of a triangular mesh approximating the vesicle shape
\begin{equation}
    E_{\rm b}=\frac B2\sum_\alpha^\text{vertex} (2 H_\alpha)^2 ~ \Delta A_\alpha
    \label{eq: SE}
\end{equation}
where $\alpha$ is the vertex label.  Here, $\Delta A_\alpha$ is the effective area corresponding to the $1/3$ of the area of triangular faces that meet at $\alpha$, and $H_\alpha$ is the discrete approximation of the mean curvature at $\alpha$ derived from the normal gradient of $\Delta A_\alpha$ \cite{meyer2003discrete,crane2018discrete}.  Full details for the implementation of the bending energy minimization in Surface Evolver are provided in Appendix~\ref{app: SE}.  Briefly, the workflow is summarized in Fig.~\ref{fig: methods}B.  Beginning with a triangular mesh (of order $10^4$ vertices), an upper section is flattened to a planar disk, of a size that targets the appropriate planar fraction $\Phi$.  Vertices of the planar inclusion are held fixed while eq.~(\ref{eq: SE}) is minimized subject to target values of enclosed volume and fluid area.

Like the solutions to the shape equation, resulting equilibria are analyzed below in terms of shape (curvature) and energetics (see example on the right half of Fig.~\ref{fig: methods}C). As the equilibria of this finite element model are not required to have axisymmetry, we also analyze numerical ground states according to the following symmetry metrics.  First, we characterize the gyration tensor defined as
\begin{equation}
S_{ij}=A^{-1} \sum_\alpha^\text{vertex} ({\bf x}_{\alpha})_i ({\bf x}_{\alpha})_j ~ \Delta A_\alpha
    \label{eq: gyration}
\end{equation}
where ${\bf x}_{\alpha}$ is the position of vertex $\alpha$ relative to the centroid of the vesicle, and $i , j = x, y$ or $z$ are Cartesian directions.  The eigenvalues, $S_1 \geq S_2 \geq S_3$, of the gyration tensor correspond to the principle moments of the mass distribution, and the corresponding orthonormal eigenvectors $\{{\bf e}_1, {\bf e}_2, {\bf e}_3 \}$, which are the principle axes of the vesicle shape.  Following these definitions, the respective ${\bf e}_1$ and ${\bf e}_3$ directions characterize the maximal and minimal ``width'' axes of the vesicle, and the trace $S_{ii} = S_1+S_2+S_3 = R_{\rm g}^2$ gives the square {\it radius of gyration} of its shape.  With the exception of the $\Phi \to 0$, ${\bar v} \to 1$ spherical limit, all shapes in this study correspond to unequal principle moments.  Based on the dissimilarity of mass distribution along orthogonal (principle) axis we classify shapes as either {\it oblate} if $S_1-S_2<0.1 R_{\rm g}^2$, as {\it prolate} if $S_2-S_3<0.07 R_{\rm g}^2$, or otherwise, as {\it triaxial}.

Additionally, we characterize the symmetry of ground state shapes in terms of the {\it centering} of the planar inclusion on the ground state vesicle.  Defining ${\bf X}_{\rm planar}$ as the vector separating the center of the inclusion from the centroid of the vesicle, we define a configuration as {\it centered} if ${\bf X}_{\rm planar}$ lies along a principle axes of the  vesicle shape.  We define alignment in terms of the angle $\omega$ between ${\bf X}_{\rm planar}$  and the nearest principle axis, defined by
\begin{equation}
\cos \omega = {\rm max}_a |\hat{{\bf X}}_{\rm planar}\cdot {\bf e}_a|,
\end{equation}
where $a=1, 2$ or 3. We categorize the shapes as {\it centered} if $\omega < 0.01 $ and {\it off-centered} otherwise.

\begin{figure*}[t!]
\begin{center}
\includegraphics[width=0.7\textwidth]{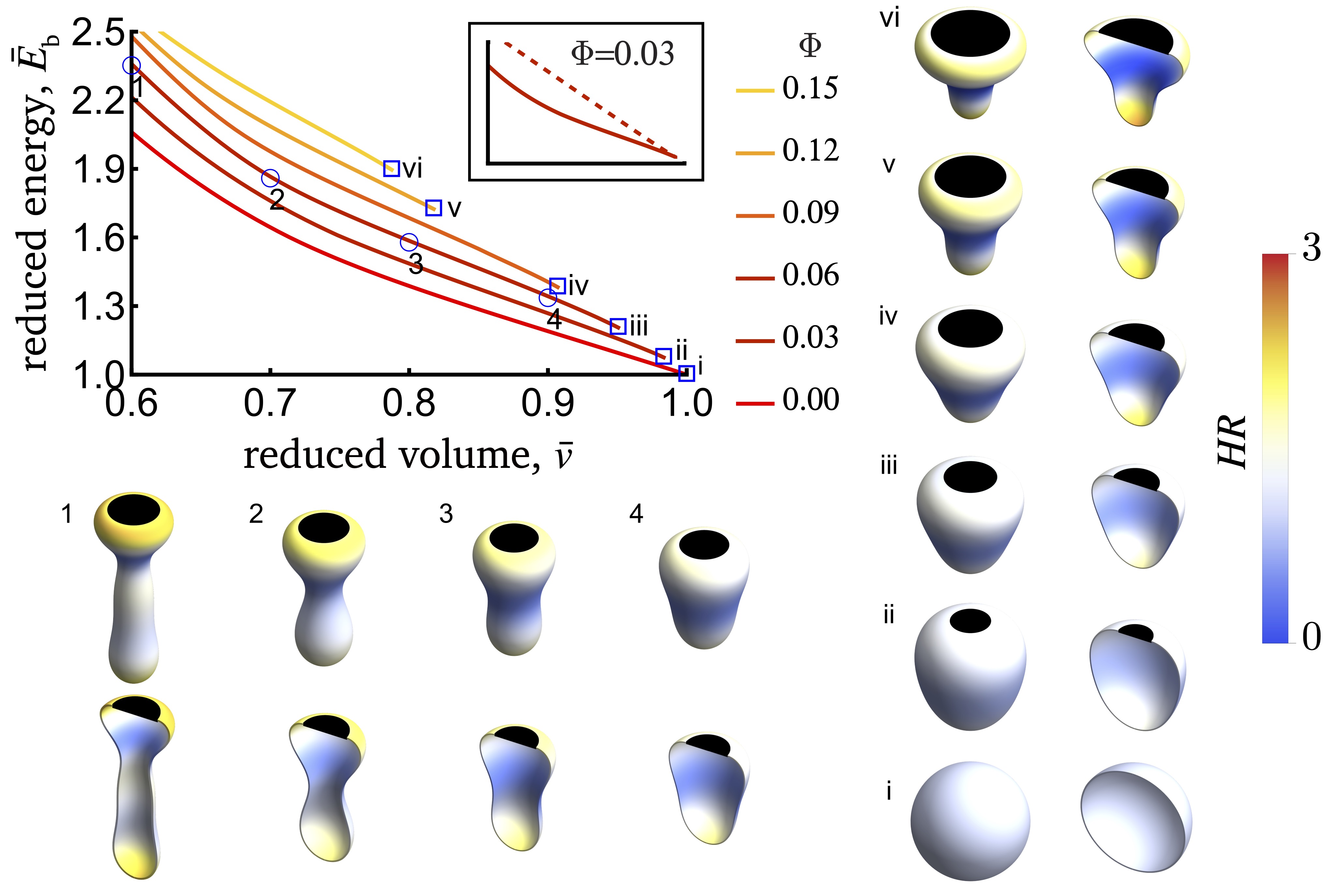}%
\end{center}
\caption{\label{fig: prolate}Axisymmetric prolate shape equilibria.  (A) Plot of reduced energy vs. reduced volume of prolate solutions for a sequence of increasing planar fractions:  $\Phi=0$, 0.03, 0.06, 0.09, 0.12, 0.15.    The inset shows the full 1D branch, which includes the lower energy branch (solid) and the higher energy branch (dashed) that merge at the maximal reduced volume of $\bar{v} = 0.983$.  (B) Shows the evolution of prolate equilibria for fixed planar fraction $\Phi=0.06$, with fluid surfaces colored according to the scaled mean curvature ($H R$) and cross-sections highlighting the detailed shapes.  (C) Show sequence of the maximal reduced volume prolate equilibria for a sequence of $\Phi$ values, illustrated in the same style as (B).}
\end{figure*}

\section{Axisymmetric shape equilibria}

\label{sec: axisymmetry}

We first describe the axisymmetric equilibria of the fluid-planar inclusion composite vesicle.  In particular, we focus on the evolution of oblate and prolate families of solutions with increasing planar fraction and on the axisymmetric ground state energy landscape as a function of $\Phi$ and $\bar{v}$.

%\subsection{Two solution branches; prolate and oblate}

% \begin{figure}[t!]
% \begin{center}
% \includegraphics[width=0.5\columnwidth]{homogeneous.jpg}%
% \end{center}
% \caption{\label{homogeneous}Energetics of two solutions branches of homogeneous vesicles; prolate (red) and oblate (blue) within the reduced volume range $0.6\le\bar v\le1$. $\bar v=1$ corresponds to a sphere which is indeed the lowest energy configuration for both branches\cite{seifert1991shape}.}
% \end{figure}

%Under the same axisymmetric setup, the shapes of homogeneous vesicles ($\Phi=0$) were studied intensely in\cite{seifert1991shape}. The elastic energy is rescaled by $8\pi B$;
%\begin{equation}
%    \bar E_\text{b}=\frac{E_\text{b}}{8\pi B}
%\end{equation}
%and we used the same rescaling in this paper. There are other solution branches with higher energy as well, but in this paper, we focused on two solution branches; prolate and oblate which are ground-state solutions for the reduced volume range $0.6\le\bar v\le 1$. Within the range, the oblate branch is favored when $0.6\le\bar v<0.65$, the prolate branch is favored when $0.65<\bar v<1$, and the two branches merge at $\bar v=1$ (sphere).

\subsection{Prolate branch}

Prolate branches of axisymmetric solutions are shown for a series of planar fractions $(\Phi)$ in Fig.~\ref{fig: prolate}.  In practice, solution branches are generated via the solution of the ODEs described in Appendix~\ref{app: axi}.  Solution branches at fixed $\Phi$ are swept out by incrementing parameters (e.g. curvature at the lower pole of the vesicle) leading to a 1D family of solutions.  As shown in the inset of Fig.~\ref{fig: prolate}A, the 1D family extends along a lower branch (solid curve) up to an upper limit of $\bar{v}$, and then folds back to a higher energy branch (dashed curve) that extends to a minimal value of  $\bar{v}$, below which no solution could be found.  For our purposes, we describe only this lower energy branch as it competes with oblate shapes in the ground state phase diagram.

Focusing first on the reduced energy $\bar E_\text{b}$ Fig.~\ref{fig: prolate}A, we note two principle effects of increasing solid fraction from the homogeneous case at $\Phi = 0 $.  First, we observe that the maximal value of $\bar{v}$ for prolate equilibria decreases with increasing $\Phi$ (see limiting solutions in Fig.~\ref{fig: prolate}C). Second, we note that the reduced energy increases with $\Phi$, i.e. at fixed $\bar{v}$.  This latter effect can be intuitively understood by inspection of the equilibria at fixed $\Phi=0.03$ in Fig.~\ref{fig: prolate}B.  Axisymmetry limits the planar inclusion to the pole of prolate shapes, which in the absence of the inclusion (i.e. $\Phi =0$) would otherwise be the location of maximal Gaussian curvature.  The inclusion of a finite-size planar inclusion at the pole of the vesicle then tends to require Gaussian curvature to be redistributed towards the opposing pole, breaking the otherwise apolar symmetry of the prolate shapes.  As $\Phi$ increases, the ``polar'' asymmetry in the curvature distribution of prolate shape grows, evidently leading to increase in the overall bending energy of the vesicles.  We contrast this effect of shape frustration on prolate vesicles with the oblate case below.

%We then investigated how the two branches change as increasing the area fraction, $\Phi$. Figure~\ref{prolate} shows the energetics and example shapes/cross-sections of the prolate branch. The Arabic numbers follow $\Phi=0.06$, and the Latin numbers follow the lowest $\bar E_\text{b}$ value of each $\Phi$. The shapes were colored by their scaled mean curvature $\bar H$ distribution
% \begin{equation}
%     \bar H=HR
% \end{equation}
%where $R=\sqrt{A/4\pi}$ is size of the vesicle. The overall trends are not much different from that of the homogeneous vesicles, but the energy gets shifted up-left as $\Phi$ gets bigger. Also from the example shapes, we can see the bending energy is focusing around the bottom part as $\bar v$ increases.

\begin{figure*}[t!]
\begin{center}
\includegraphics[width=0.7\textwidth]{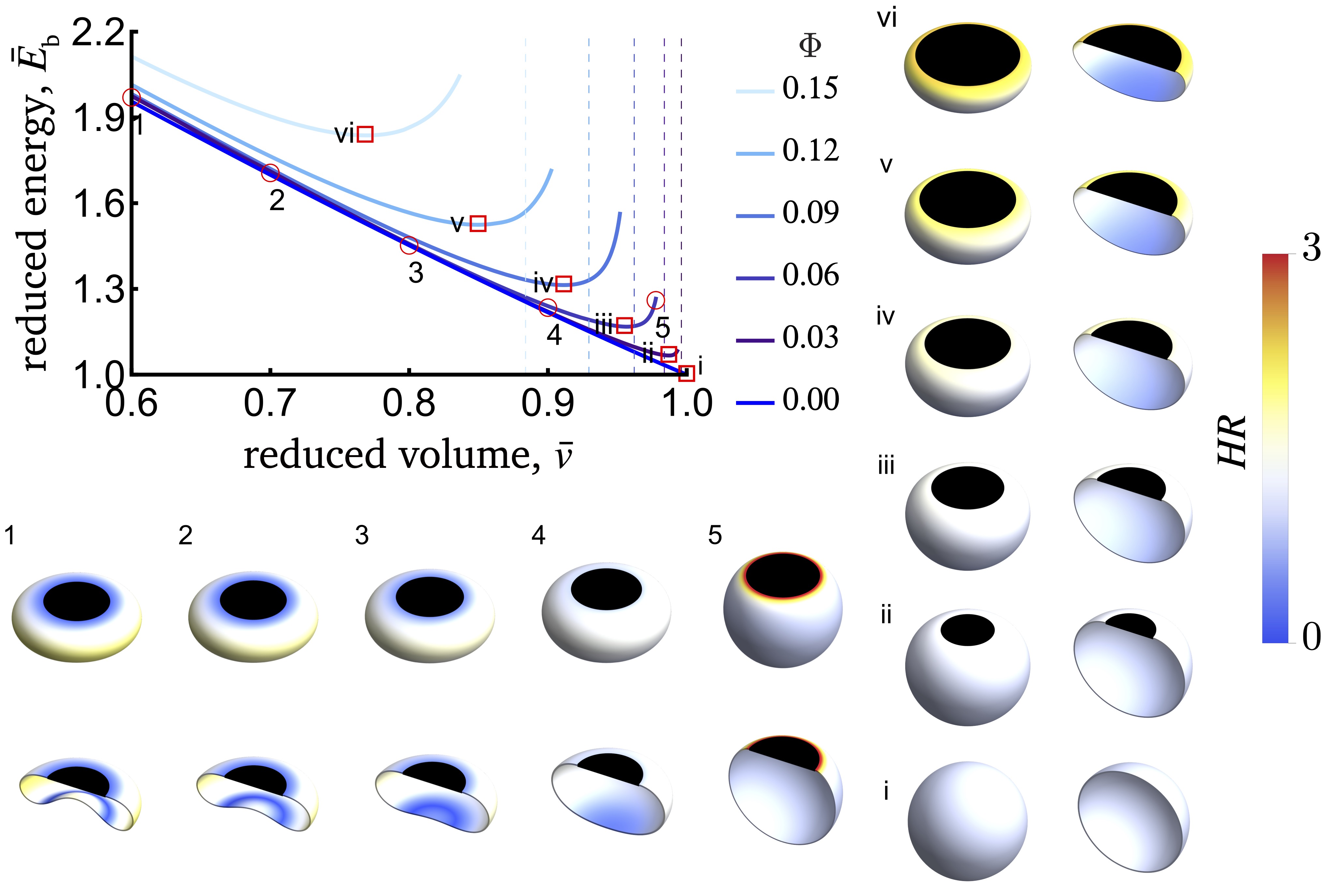}%%
\end{center}
\caption{\label{fig: oblate}Axisymmetric oblate shape equilibria.  (A) Plot of reduced energy vs. reduced volume of prolate solutions for a sequence of increasing planar fractions:  $\Phi=0$, 0.03, 0.06, 0.09, 0.12, 0.15.  The dashed vertical lines indicate the location of the corresponding isoperimetric limiting inflation values $\bar{v}_{\rm max}(\Phi)$.  (B) Shows the evolution of oblate equilibria for fixed planar fraction $\Phi=0.1$, with fluid surfaces colored according to the scaled mean curvature ($H R$) and cross-sections highlighting the detailed shapes.  (C) Show sequence of the maximal reduced volume oblate equilibria for a sequence of $\Phi$ values, illustrated in the same style as (B).}
\end{figure*}

\begin{figure}[t!]
\begin{center}
\includegraphics[width=\columnwidth]{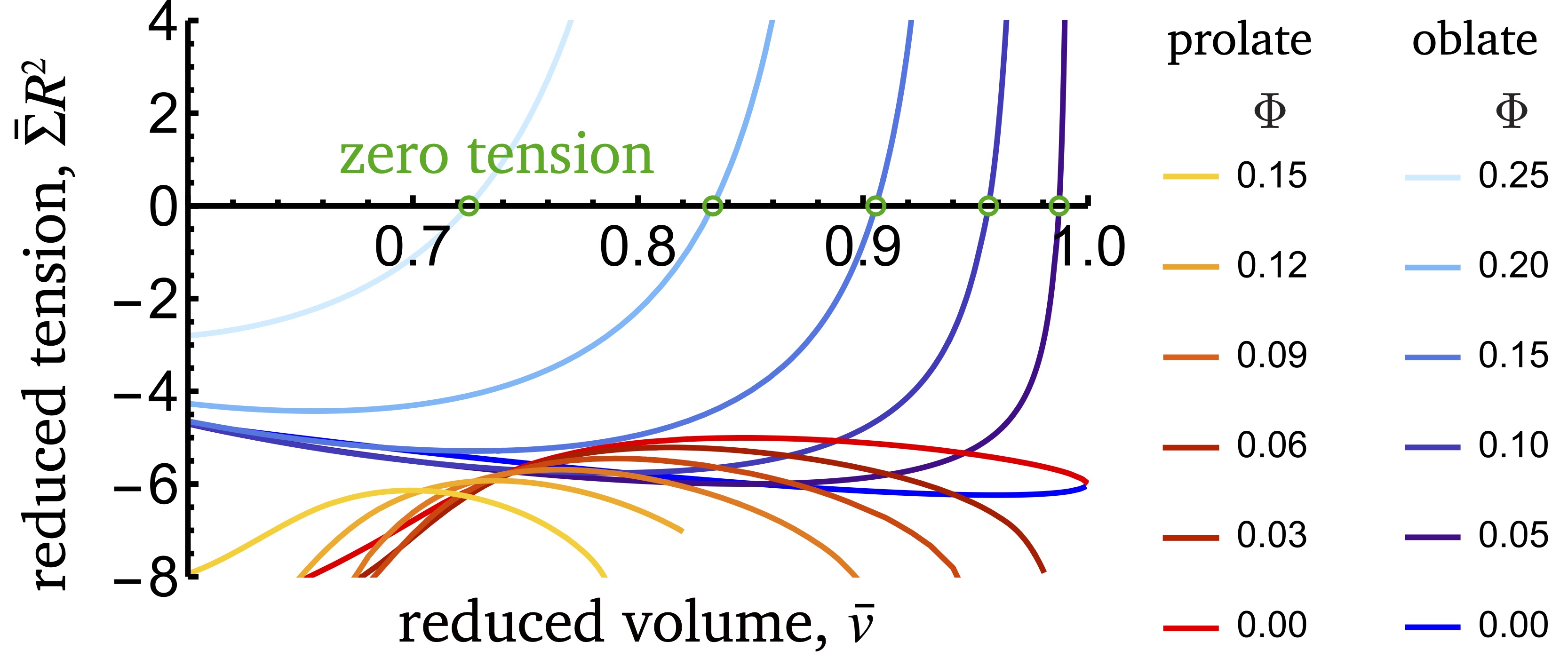}%
\caption{\label{fig: tension}Plots of reduced tension versus reduced volume for oblate (blue curves) and prolate (red/orange curves) solution branches of axisymmetric solutions for a sequence of planar fractions indicated in the legend. }
\end{center}
\end{figure}

\subsection{Oblate branch}

In Fig.~\ref{fig: oblate} we show the structure and energetics of the oblate branch of axisymmetric shape equilibria.  As for the prolate branch, we plot the energy dependence on reduced volume for a sequence of fixed planar fraction $\Phi$ in Fig.~\ref{fig: oblate}A as solid curves.  The dashed vertical lines indicate the predicted location of the isoperimetric limit volumes $\bar{v}_{\rm max} (\Phi)$.  We note for the oblate branch, solutions of the shape equations become very stiff in the large $\bar{v}$ regime, leading to a failure of our numerical method to resolve solutions beyond a certain upper limit of $\bar{v}$ for each $\Phi$ which falls below $\bar{v}_{\rm max} (\Phi)$.  Unlike the prolate case, however, we discuss in Sec. \ref{sec: nonaxi} the oblate shape equilibria exist in this regime that extends fully to the asymptotic limit $\bar{v} \to \bar{v}_{\rm max} (\Phi)$.

Like the prolate solutions, increasing solid fraction increases the bending energy, shifting curves $\bar{E}_{\rm b} ( \bar{v})$ upward with increasing $\Phi$.  Again, this increase in energy can be attributed to the necessary redistribution of Gaussian curvature imposed by the planar inclusion.  However, it may be intuitive to expect that this effect could be less geometrically disruptive, and therefore less energetically costly, for oblate shapes since the planar inclusion occupies a polar region which is the relatively flatter portion of the oblate equilibria.  We describe the consequences of this for the ground state shape phase diagram below.

The most notable difference between the oblate and prolate branches is the non-monotonic dependence of bending energy on reduced volume shown in Fig.~\ref{fig: oblate}.  For low $\bar{v}$, the energy of oblate shapes decreases with increasing inflation consistent with the homogeneous ($\Phi =0$) and prolate case, until a point $\bar{v}^*(\Phi)$ where $\bar{E}_{\rm b} $ is minimal with respect to inflation.  At higher inflation, $\bar{v} >\bar{v}^*(\Phi) $ the bending energy increases with $\bar{v}$. As we discuss below in Sec. \ref{sec: nearlyiso}, the structure and energetics of the high inflation limit derive from the concentration of bending energy at the fluid-planar boundary that matches the planar inclusion shape to the fluid membrane, where the fluid membrane shape becomes increasingly spherical as inflation increases.  This non-monotonic form of $\bar{E}_{\rm b}$ also implies the existence of an {\it optimal inflation} state (i.e. minimal bending energy) for a given planar fraction at $\bar{v}^*(\Phi)<1$, in clear contrast with the homogeneous vesicle which is minimal in the spherical limit $\bar{v} \to 1$.

According to eqs.~(\ref{eq: ten}) and (\ref{eq: press})  the non-monotonic dependence of $\bar{E}_{\rm b}$ on reduced volume has implication for the equation of state relating membrane tension $\bar{\Sigma}$ to the inflation of composite vesicles.  Fig. \ref{fig: tension} compares the predicted values of reduced tension as functions of reduced volume between oblate (blue) and prolate (orange/red) axisymmetric shapes.  Notably, tension (and pressure) is always negative for prolate shapes, indicating that these are in effect always below optimal inflation at all $\bar{v}$.  In contrast, oblate shapes exhibit both {\it under-inflation} at low reduced volumes (i.e. $\bar{\Sigma} (\bar{v} <\bar{v}^*)<0$) and {\it over-inflation} at higher reduced volumes (i.e. $\bar{\Sigma} (\bar{v} >\bar{v}^*)>0$) .  As the elastic energy is minimal at $\bar{v}^*$ for fixed $\Phi$, these states correspond to vanishing tension.  In the remainder of this article, we refer to the conditions $\bar{v}^*(\Phi)$ as {\it zero-tension} states, which notably belong only to the oblate family.

%\subsubsection{Optimal inflation; zero tension states}

%One useful piece of information about the tension is the relation with the bending energy. By taking $\delta F=0$ from eq.~(\ref{freeenergy}), we get the following thermodynamic relation
%\begin{equation}\label{tensioneq}
%    A\bar\Sigma=12\pi\bar v\frac{\partial \bar E}{\partial \bar v}
%\end{equation}
%This indicates that tension is the rate of the bending energy to the reduced volume. The tension is negative if $\bar E(\bar v)$ plot has a negative slope, positive if $\bar E(\bar v)$ plot has a positive slope, and tension-free if the slope is zero.

%The existence of $\bar v^*$ also implies the existence of the tension-free state and positive tension state which doesn't show up for the homogeneous vesicles (and the prolate branch for $\Phi>0$ as well). From eq.~(\ref{tensioneq}), $\bar v^*$ is indeed the tension-free state, and the tension becomes negative at $\bar v<\bar v^*$ and positive at $\bar v>\bar v^*$ [See Figure~\ref{tension}], or equivalently, we can tell the shapes with $\bar v<\bar v^*$ are deflated and shapes with $\bar v^*$ are inflated.

\begin{figure*}[t!]
\begin{center}
\includegraphics[width=\textwidth]{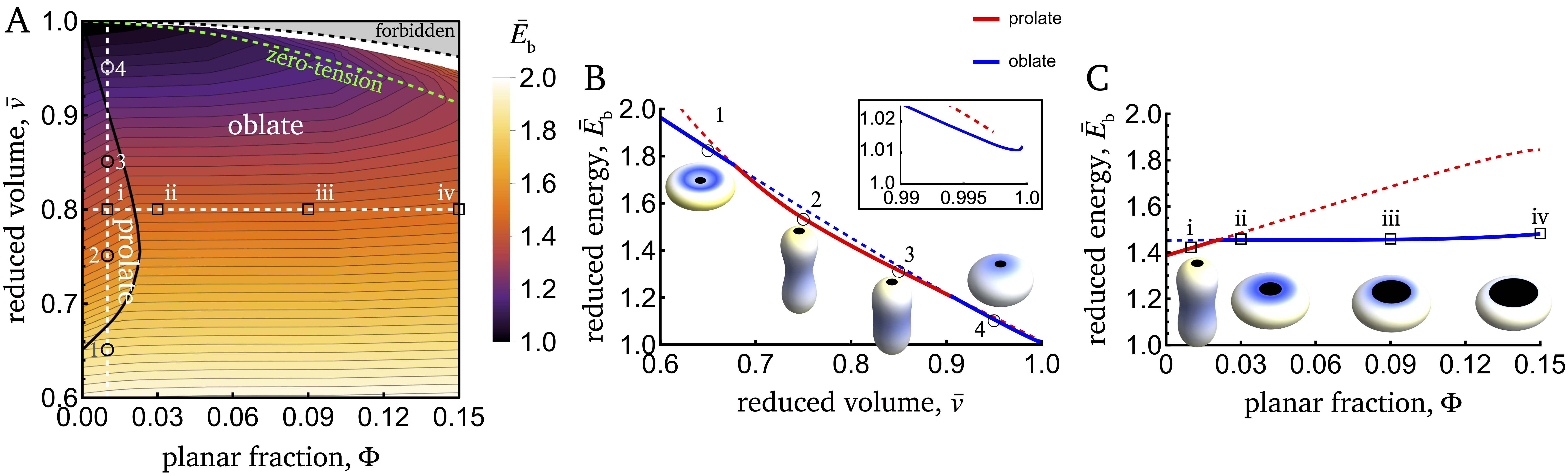}%
\end{center}
\caption{\label{fig: phasediagram} (A) Contour plot of the reduced energy of the minimal energy axisymmetric shapes as a function or reduced volume $\bar{v}$ and planar fraction $\Phi$.  Regions are highlighted according to symmetry of lowest energy shape:  {\it prolate} for low $\Phi$ and intermediate $\bar{v}$; and {\it oblate} elsewhere.  Colored regions correspond to regions where axisymmetric solutions are identified.  In white regions, near the isoperimetric limit, numerical solutions could not be obtained.  (B) Plot of re-entrant oblate-prolate-oblate shape along the vertical sequence (increasing reduced volume) as indicated in (A).  (C) Plot of prolate-oblate transition at fixed-$\bar{v}$ and increasing $\Phi$ sequence as indicated in (A).}
\end{figure*}

\begin{figure}[t!]
\begin{center}
\includegraphics[width=0.85\columnwidth]{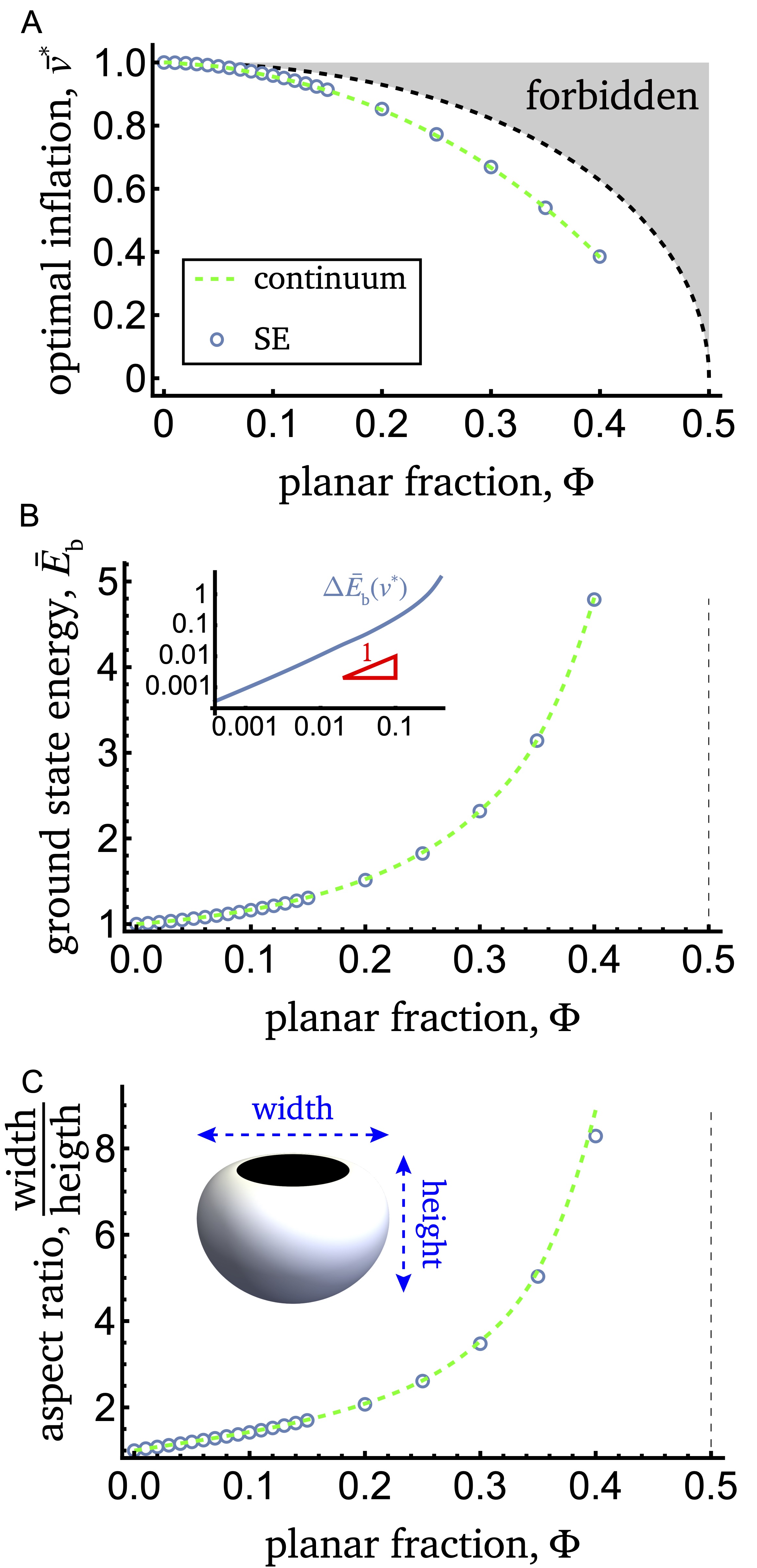}%
\end{center}
\caption{\label{fig: optimal}(A) optimal reduced volume $\bar v^*$ as a function of area fraction $\Phi$. (B) ground state energy at the optimal reduced volume. The inset is log-scaled plot of $\Delta \bar E_\text{b}(\bar v^*)=\bar E_\text{b}(\bar v^*)-1$ as a function of $\Phi$. $\Delta \bar E_\text{b}(\bar v^*)\approx\Phi$ for small $\Phi$.}
\end{figure}

\subsection{Phase diagram of axisymmetric ground state shapes}

In Fig.~\ref{fig: phasediagram} we present the axisymmetric ground state shapes as a function of reduced volume and planar fraction.  As shown in Figs.~\ref{fig: prolate} and \ref{fig: oblate}, relative to the homogeneous case, incorporating a planar region increases the elastic energies of both prolate and oblate solution branches.  However, $\bar{E}_{\rm b}$ increases more for given $\Phi$ for {\it prolate} shapes since the inclusion occupies the region of otherwise high curvature that has to be redistributed to the opposite (fluid) pole of the vesicle.  Hence, for small planar fraction $0<\Phi<\Phi_{c}\simeq 0.023$, the vesicle shape exhibits a ``re-entrant'' dependence on increasing $\bar{v}$ as shown in Fig.~\ref{fig: phasediagram}B, with oblate shapes stable in the low and high inflation regions separated by an intermediate window of stable prolate shapes.  As $\Phi$ increases, the window of intermediate stable prolate shapes shrinks until it completely disappears at a {\it critical point} $\Phi_{c}\simeq 0.023$ and $\bar{v}_{c}\simeq 0.76$.  Thus, even an inclusion of a relatively small size has a surprisingly strong effect on the shape and energetics of axisymmetric states.  Outside of this region, oblate shapes are generically favored.  In particular, this implies that for fixed $\Phi$ minimal energy states are always oblate and fall along the zero-tension line (shown in green in  Fig.~\ref{fig: phasediagram}A).

In Fig.~\ref{fig: phasediagram}C, we plot the analogous 1D cut for fixed $\bar{v}$ and increasing $\Phi$, which shows that $\bar{E}_{\rm b}$ is an increasing function of $\Phi$, marked by a transition between a steep energy dependence for prolates to a weaker, but still increasing, dependence for oblates.  We note that a similar prolate to oblate transition was reported for a model 2-phase vesicles~\cite{Gutlederer_2009}, with increasing contrast in bending stiffness, presumably driven by the cost of locating the stiffer/flatter domain at the high-curvature pole.  

Hence, while it was shown for homogeneous vesicles that oblate solutions occupy a relatively narrow regime of the parameter space for low reduced volume~\cite{seifert1991shape}, a primary conclusion of the axisymmetric shape phase diagram in Fig.~\ref{fig: phasediagram}A, is that the incorporation of planar inclusions, even at low area fractions, expands the stability of oblate over prolate shapes over much of the 2-dimensional parameter space.  This is notably the case for the higher reduced volume regime and the minimal-energy shapes $\bar{v}_*(\Phi)$.  Fig.~\ref{fig: optimal} analyzes the energetics and shape of these minimal-energy shapes as a function of $\Phi$, and also compares axisymmetric solutions to the shape equation to results of the Surface Evolver simulations (shown as blue circles). While axisymmetry is not imposed for Surface Evolver simulations, we show below (in Sec.~\ref{sec: nonaxi}) that solutions retain axisymmetry in the regimes occupied by the zero-tension states shown in Fig.~\ref{fig: optimal}A, which track but fall somewhat below the isoperimetric limit shapes in the $\Phi$-$\bar{v}$ plane.   Fig.~\ref{fig: optimal}B shows the reduced energy of optimal inflation (axisymmetric) shapes as a function of the planar fraction.  Notably, bending energy increases monotonically with $\Phi$ from the homogeneous case of $\bar{E}_{\rm b}(\Phi =0) = 1$, growing with linear dependence in small-$\Phi$ limit.  This linear slope and super-linear behavior at larger $\Phi$ are consistent with the previously studied critical adhesion transition behavior for homogeneous vesicles on planar substrates (at zero pressure), in which the size of adhered contact zone grows continuously from zero above a critical adhesion strength~\cite{seifert1991adhesion,lipowsky1991adhesion}.  In Fig.~\ref{fig: optimal}C, we plot the width to height aspect ratio of zero-tension vesicles as a function of planar fraction, which shows that even fairly small inclusions of a few $\%$ area fraction strongly warp the ground state shape away from the spherically symmetric homogeneous case.  Notably, for homogeneous fluid vesicles such highly oblate anisotropies may typically require
non-trivial experimental conditions due to the high degree of pressure necessary for those states.  As the incorporation of planar inclusion shifts the symmetry of the zero-tension state considerably, these otherwise ``underinflated'' highly oblate shapes indeed become stable and long-lived configurations in composite vesicles.

%It was found that the prolate branch is mostly favored in the reduced volume range for the homogeneous vesicles\cite{seifert1991shape}[See Figure~\ref{homogeneous}]. Now we can add another axis about the area fraction and see which branch is favored over the other in the 2-dimensional parameter space. Figure~\ref{phasediagram} shows the phase diagram and example phase transitions along the two axes (burgundy, olive). Unlike the homogeneous vesicles, the oblate branch is occupying more and more range of reduced volume as the area fraction increases. Along the phase boundary (black), the two branches are degenerate. The prolate branch can be favored for some range of reduced volume for up to $\Phi=\Phi_\text{critical}\approx0.023$, but for $\Phi>\Phi_\text{critical}$, the oblate branch is always preferred. The optimal inflation $\bar v^*(\Phi)$ follows the green dashed curve in Figure~\ref{phasediagram}. The black dashed curve denotes $\bar v_\text{max}$, and the solutions are theoretically upper bound by the curve. The white region is where the solutions are analytically unbounded but numerically suppressed.

\begin{figure*}[t!]
\begin{center}
\includegraphics[width=0.75 \textwidth]{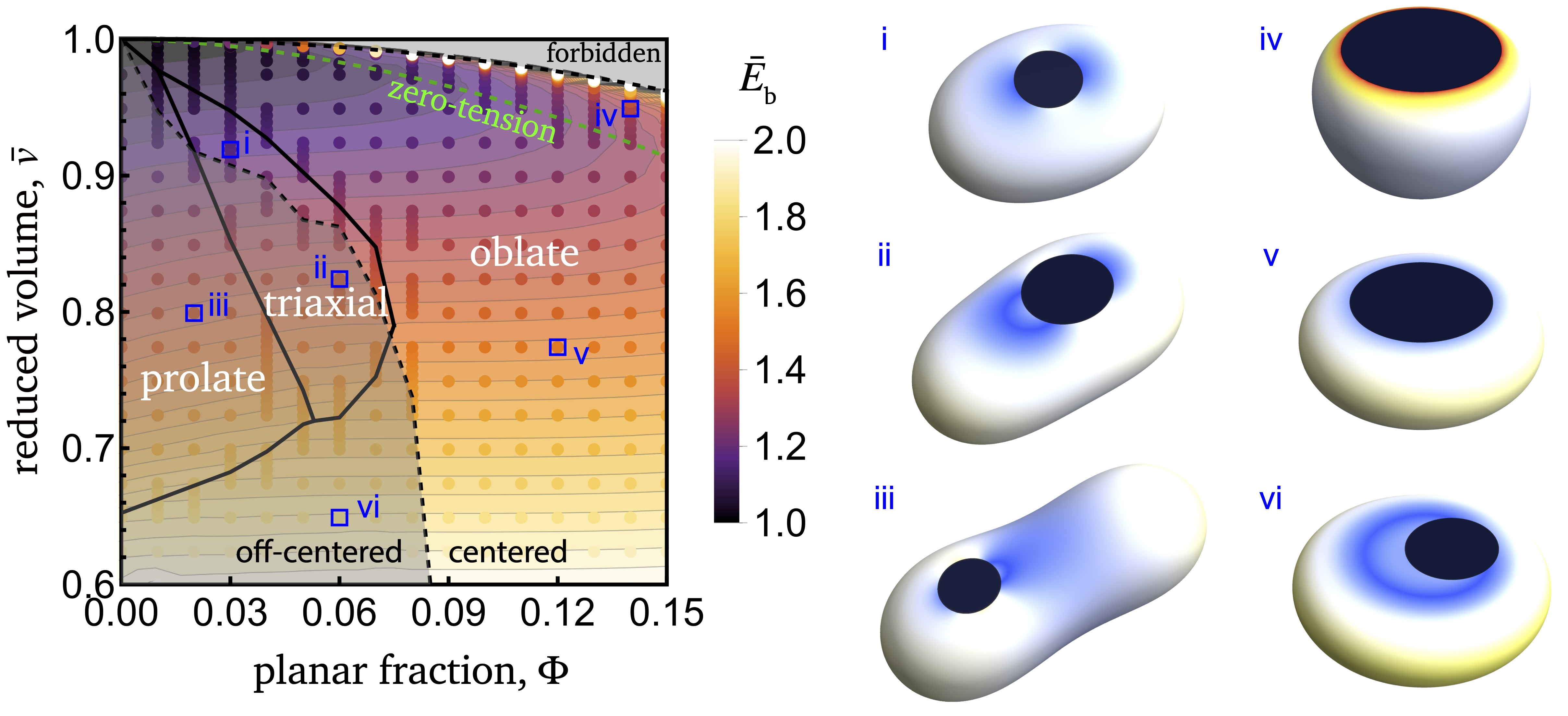}%
\end{center}
\caption{\label{fig: SEphase}Phase diagram of shape equilibria and example configurations. The shape equilibria separate into prolate, triaxial, and oblate branches. The domain position is off-centered on the left side of the dashed line crossing the left top corner and the middle of the bottom axis, and centered on the right side.}
\end{figure*}

\section{Axisymmetric and non-axisymmetric ground states}

\label{sec: nonaxi}

% \subsection{Symmetry breaking; non-axisymmetric equilibria}

Fig.~\ref{fig: SEphase} shows the results from Surface Evolver simulations of ground states of the fluid-planar inclusion composite vesicles in the same $\Phi$-$\bar{v}$ plane illustrated in axisymmetric shape phase diagram in Fig.~\ref{fig: phasediagram}A. Here, the simulated parameters are indicated as filled circles colored according to the reduced energy, and examples of resulting ground state shapes are shown for indicated points on the right.  

The absence of imposed axisymmetry reveals that ground state shapes break axisymmetry over some, but not all regions of the parameter space.  Foremost, we note that the location of planar inclusion breaks axisymmetry, particularly at low $\Phi$ values where the gross shape remains largely prolate.  As discussed above, constraining the planar inclusion to the high-curvature pole of prolate shapes is energetically costly. Instead, we find that prolate ground states displace the inclusion off-axis to the lower-curvature ``sides'' of the vesicle, therefore requiring less disruption of the otherwise elastically preferred prolate shape.  Notably, this off-axis shape allows for the expansion of the stable window of prolate shapes to larger $\Phi$ values, relative to the modest range possible for axisymmetric shapes.  However, as the size of the planar inclusion grows, we observe a tendency to flatten the side of the vesicle it occupies, which in turn gives rise to {\it triaxial} ground states in which the size distribution along all three principal axes is unequal (according to the criteria defined in Sec.~\ref{sec: SE}) and the inclusion plane sits normal to the thinnest direction of the vesicle.  As the size of the inclusion grows further, the dimensions of vesicle in the two directions perpendicular to the inclusion plane become more and more similar, eventually becoming equal as the ground state transitions to an oblate shape where the inclusion is centered at the flat, polar region.  

Taken together, these observations show that breaking of axisymmetry alters the shape phase diagram at low $\Phi$ by displacement of the inclusion to locally flatter regions of the global shape, leading to an expansion of the non-oblate shape region to large planar fractions, up to about $\Phi\approx 0.07$--$0.08$.  However, we find that ground states retain axisymmetric oblate shapes in regimes of sufficiently high inflation and planar fractions.  This is notably the case at and near the zero-tension regions analyzed in Fig.~\ref{fig: optimal}.  We take advantage of the emergent axisymmetry of oblate ground states in this high inflation regime to analyze the shape space intermediate to the zero-tension states and the isoperimetric limit shapes in the following section.

% Despite the good agreement on the shape examples in Figure~\ref{SEplot}, there are still possibilities of symmetry breaking. As we conjectured in Section~\ref{continuumsection}, the energy gain due to the planar inclusion is closely related to how much deformation is required to flatten out a portion of the surface. In that sense, since prolates have larger Gaussian curvature at the top and the bottom while relatively smaller Gaussian curvature over the side, we can imagine putting a planar inclusion on the side surface instead to introduce less deformation breaking the axisymmetry. This scenario should appear more easily especially if the area fraction is small. However, as the reduced volume approaches $\bar v_\text{max}$ the shape must recover the axisymmetry.

%We classified the simulation shapes according to the shape category we defined in Section~\ref{momentofinertiasection}. Figure~\ref{SEphase} shows the phase diagram and example phase transitions along the two axis directions. Along the burgundy arrow, the phase changes thrice, from symmetry broken oblate to symmetry broken prolate, then to symmetry broken oblate again, and then to axisymmetric oblate. Likewise, along the olive arrow, the phase changes twice from symmetry broken prolate to symmetry broken oblate, and then to axisymmetric oblate.

\section{Curvature concentration in nearly isoperimetric regime}

\begin{figure*}[t!]
\begin{center}
\includegraphics[width=\textwidth]{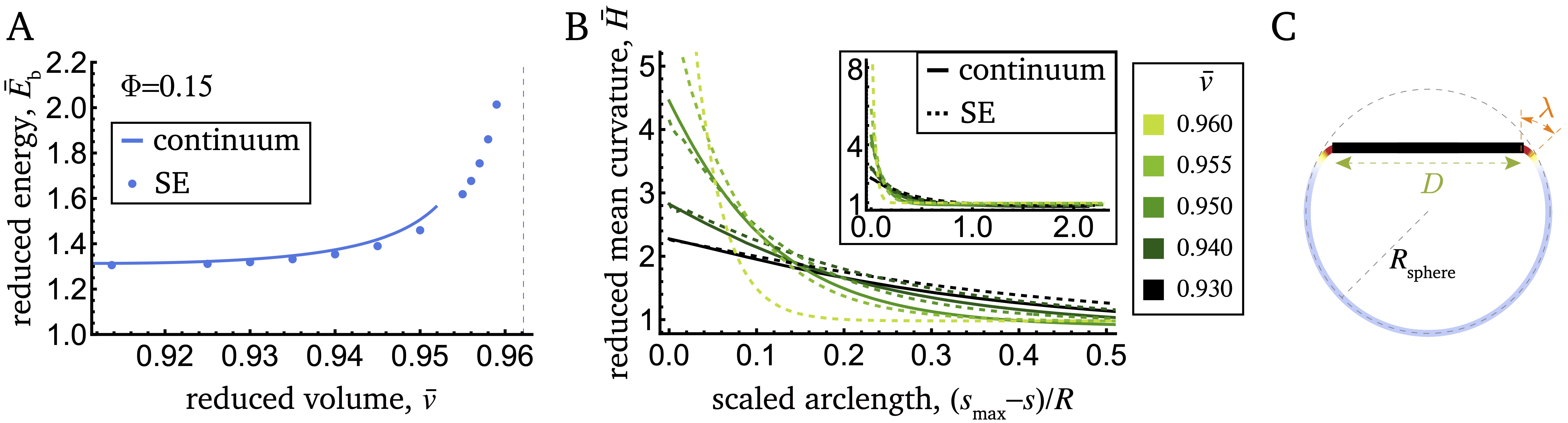}%
\end{center}
\caption{\label{fig: nrlyiso}Shape behavior at $\bar v\lesssim\bar v_\text{max}$. (A) bending energy for $\Phi=0.15$. (B) curvature distribution along the arclength. The inset is for the full range of arclength (C) cross-section of $\bar v=0.952 $ overlayed on a circle. $R_\text{sphere}$ is the radius of overlaying circle, and $\lambda$ is the hinge length. $R=\sqrt{A/4\pi}$ is the size of the vesicle.}
\end{figure*}

\label{sec: nearlyiso}

Given the emergent axisymmetric oblate shapes for high reduced volume demonstrated in the previous section, here we focus on the structure and energetics of composite shapes in the high-tension regime, as vesicles approach the isoperimetric limit volume $\bar{v} \to \bar{v}_{\rm max} (\Phi)$.  Our analysis is based on characteristic features of oblate shapes in this regime, shown in Fig.~\ref{fig: nrlyiso} for a particular case of $\Phi =0.15$.  

Fig.~\ref{fig: nrlyiso}A first compares the reduced energy for $\bar{v} > \bar{v}_*$ for continuum solutions to the axisymmetric shape equations (solid curve) and Surface Evolver simulations (filled circles).  These show strong quantitative agreement, notwithstanding the fact that continuum solutions cannot be extended beyond $\bar{v} \simeq 0.953$ which falls somewhat below the isoperimetric limit of $\bar{v}_{\rm max} \simeq 0.962$.  Notably, Surface Evolver simulations are found in this intermediate regime and confirm the tendency of bending energy to grow as the isoperimetric limit is approached.  

Fig.~\ref{fig: nrlyiso}B shows the structural pattern in this high-inflation regime by plotting the reduced mean curvature $\bar{H}= HR$ as a function of arc position along the fluid contour.  These generically show that curvature is maximal at the planar inclusion edge $s=s_{\rm max}$ and then rapidly falls much lower to a nearly constant value on the bottom pole, i.e. $H(s\to 0) = 1/R_{\rm sphere} \simeq 1/R$, consistent with an approximately spherical shape away from the inclusion edge.  Defining the length scale $\lambda$ to characterize the {\it size} of the high bending region, it is notable that $\lambda$ decreases while edge curvature $H(s_{\rm max}) $ grows as $\bar{v} \to \bar{v}_{\rm max}$, indicating the curvature and bending energy become increasingly concentrated in this regime.

\begin{figure*}[t!]
\begin{center}
\includegraphics[width=1.8\columnwidth]{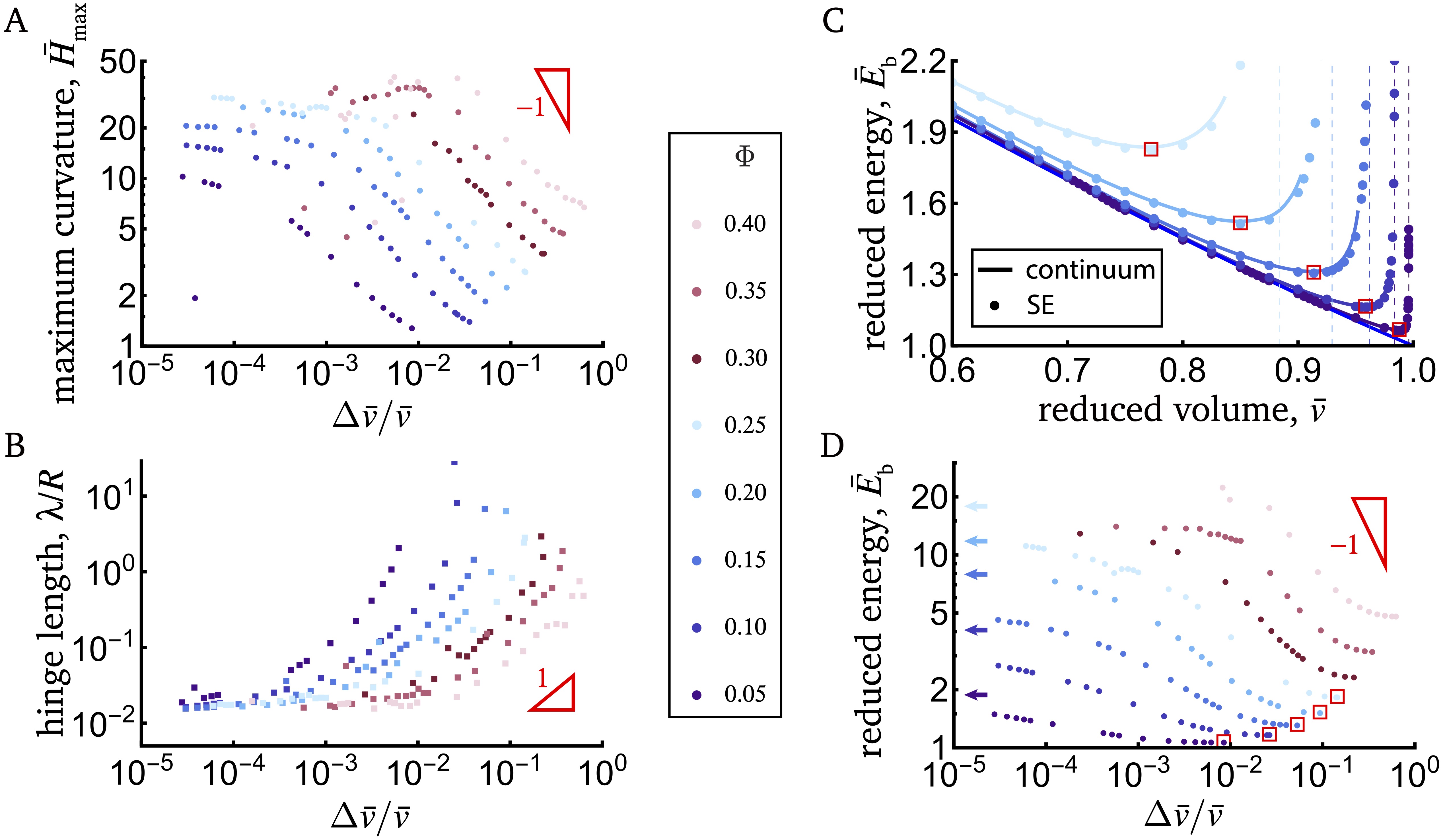}%
\end{center}
\caption{\label{fig: nrlyisoSE}For various $\Phi$, (A) log-log scaled maximum mean curvature $H_\text{max}$ as a function of $\Delta\bar v/\bar v$ (B) log-log scaled hinge length $\lambda$ as a function of $\Delta\bar v/\bar v$ (C) reduced bending energy $\bar E_\text{b}$ as a function of $\bar v$. (D) log-log scaled reduced bending energy $\bar E_\text{b}$ as a function of $\Delta\bar v/\bar v$. Red boxes ($\Box$) are the tension-free states.}
\end{figure*}

The apparent separation of scales between the high-curvature ``hinge'' at the planar edge and the nearly spherical ``bulb'' that describes the bulk shape of the fluid phase suggest the following asymptotic analysis of shape equilibrium in the nearly isoperimetric limit.  First, consideration of the shape equation, eq.~(\ref{eq: shapeeq}), in the nearly uniform spherical bulb implies the standard equilibrium condition, $P - 2 \Sigma H(s=0) \approx 0$, effectively the Young-Laplace condition
\begin{equation}
\bar{P} \approx \frac{2 \bar{\Sigma}}{R}.
\end{equation}
Near the edge, the shape of the membrane is strongly bent away from the spherical shape and exhibits a much higher principal curvature $\kappa_{\rm hinge} = \psi'(s_{\rm max}) \gg 1/R$.  Notably, for sufficiently large inclusions (i.e. $\Phi \lesssim 1$) the curvature in the hinge direction is also much larger than the other principle (hoop) direction, so we expect $H(s_{\rm max}) \approx \kappa_{\rm hinge}/2 $ and $H^2(s_{\rm max}) \gg K_{\rm G} (s_{\rm max}) $.  The membrane in the hinge region must turn through a finite angle $\Delta \psi$ to match the bulb to the planar inclusion, which can be roughly estimated as $\Delta \psi \approx \kappa_{\rm hinge} \lambda$.  We argue that this turning angle approaches a finite value corresponding to the angle discontinuity in the isoperimetric limit shape, i.e. $\Delta \psi (\bar{v} \to \bar{v}_{\rm max}) \to \theta_{\rm iso}$, which gives as an asymptotic relationship
\begin{equation}
\kappa_{\rm hinge} \approx \theta_{\rm iso}/\lambda , \ \ \ \ ({\rm for} \ \bar{v} \to \bar{v}_{\rm max})
\end{equation}
Applying this result to the shape equation, eq.~(\ref{eq: shapeeq}), in the hinge region, we expect that $\bar{P} \ll 2 \bar{\Sigma} H(s_{\rm max}) $ since $H(s_{\rm max}) \approx \kappa_{\rm hinge} \gg H(0) \simeq 1/R$, suggesting that mechanically equilibrium in the hinge results from a balance of tension and bending terms, that is
\begin{equation}
\bar{\Sigma} \approx H^2(s_{\rm max}) - \frac{\nabla_\perp^2 H(s_{\rm max})}{H(s_{\rm max})} \sim 1/\lambda^2 , \ \ \ \ ({\rm for} \ \bar{v} \to \bar{v}_{\rm max}) 
\end{equation}
where we estimated $\nabla_\perp^2 H(s_{\rm max}) \approx \kappa_{\rm hinge}/\lambda^2$ and used the scaling $\kappa_{\rm hinge}\sim 1/\lambda$.  Hence, we find that the hinge size is effectively a bendocapillary length scale $\lambda \sim \sqrt{1/\Sigma}$\cite{roman2010elasto} that becomes arbitrarily short-ranged as tension diverges in the $ \bar{v} \to \bar{v}_{\rm max}$ limit.  In this regime, elastic energy is expected to be dominated by the bending concentration in the hinge, from which we estimate the reduced energy,
\begin{equation}
 \bar E_\text{b}\approx\frac{1}{2}\kappa_\text{hinge}^2~D\lambda\sim\frac{f(\Phi) R}{\lambda}, \ \ \ \ ({\rm for} \ \bar{v} \to \bar{v}_{\rm max}) 
 \end{equation}
where $f(\Phi)$ is an ${\cal O}(1)$ dimensionless function of the planar fraction.  Because elastic energy is related to tension via thermodynamic derivative with respect to reduced volume, eq.~(\ref{eq: ten}), we have from the scaling dependence $\bar{\Sigma} \sim \lambda^{-2}$ an asymptotic relation for the dependence of the hinge size on reduced volume
\begin{equation}
\frac{d \lambda}{d \bar{v}} \sim - \frac{R}{\bar{v}}  , \ \ \ \ ({\rm for} \ \bar{v} \to \bar{v}_{\rm max}) 
\end{equation}
where we neglect, for this purpose, the $\Phi$-dependence of the prefactor.  Since we expect $\lambda \to 0$ as $\bar{v} \to \bar{v}_{\rm max}$, in the isoperimetric limit, we expect the hinge size to obey
\begin{equation}
\lambda/R \sim \ln(\bar{v}_{\rm max} / \bar{v} ) \sim \frac{\Delta \bar{v}}{\bar{v}} , \ \ \ \ ({\rm for} \ \bar{v} \to \bar{v}_{\rm max}) 
\end{equation}
where $\Delta \bar{v} = \bar{v}_{\rm max}- \bar{v}$.  Correspondingly, the expected vanishing of the hinge size suggests that hinge curvature and reduced energy diverge in the isoperimetric limit as $\kappa_{\rm hinge} \sim E_{\rm b} \sim 1/\Delta \bar{v}$.

\begin{figure*}[t!]
\begin{center}
\includegraphics[width=0.8\textwidth]{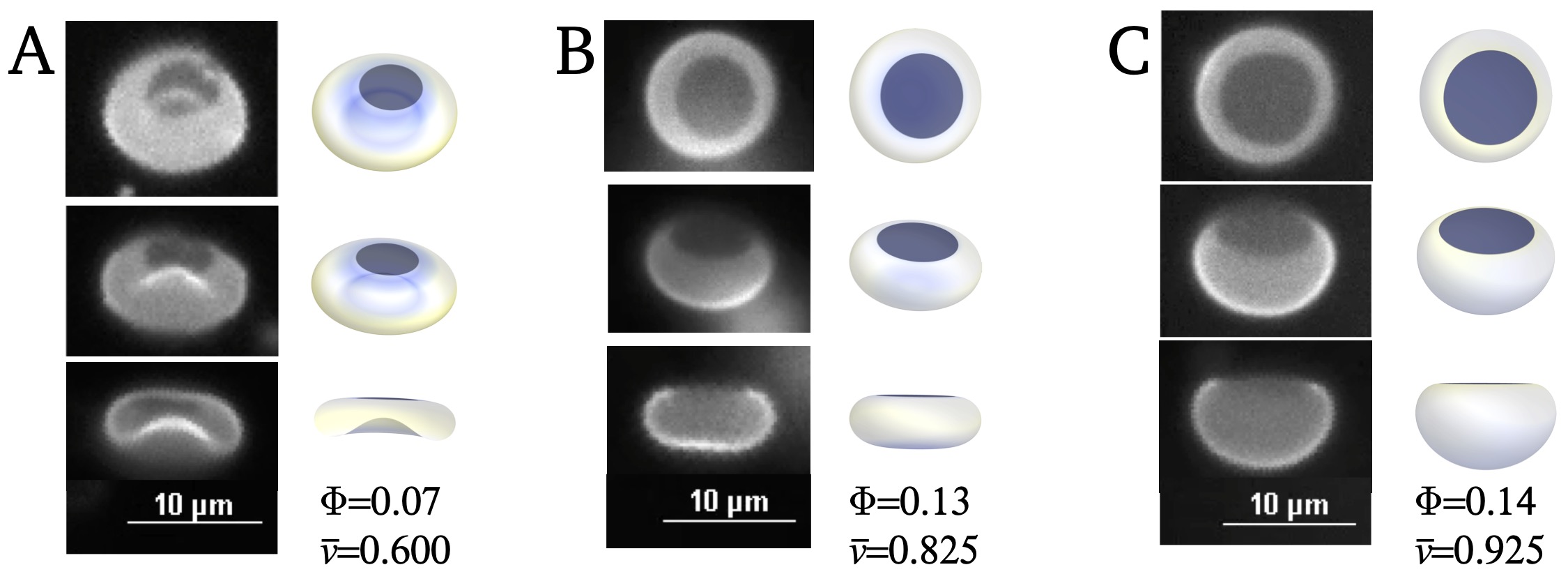}%
\end{center}
\caption{\label{fig: microscopic}Fluorescent micrographs of three fluid-solid vesicles, each shown from different views and compared to axisymmetric equilibria of fluid-planar inclusion model.  Bright regions are fluorescently labeled fluid phase (DOPC rich) while dark regions are compact 2D solid domains (DPPC rich).  Vesicle in (A) is formed from 20/80 (DPPC/DOPC) lipid composition and vesicles in (B) and (C) are formed from 30/70 compositions.  Estimated solid fractions from the images are consistent with known level-rule compositions of DPPC/DOPC mixtures cooled to room temperature~\cite{wan2023flowering}.}
\end{figure*}

In Fig.~\ref{fig: nrlyisoSE} we analyze the asymptotic approach to the isoperimetric limit inflation, based on Surface Evolver simulations, which are able to model shape relaxation further into this regime than was possible for the analytical axisymmetric solutions. Fig.~\ref{fig: nrlyisoSE}A analyzes the mean curvature at the planar edge, $H(s_{\rm max})$, and size of the hinge zone, $\lambda$, as a function of $\Delta \bar{v}$ on a log-log scale for a series of planar fractions.  We derive a measure of the hinge size from the slope of mean curvature at the planar edge, using the definition $ \frac{d H}{d s}\Big|_{s_{\rm max}} /H(s_{\rm max}) \equiv \lambda^{-1}$,  as shown in Fig.~\ref{fig: nrlyisoSE}B, extracted from fit profiles of the radial profile from 3D relaxed mesh configurations.  We note $H(s_{\rm max})$ and $\lambda$ exhibit the respective tendencies to diverge and vanish as the isoperimetric limit of $\Delta \bar{v} \to 0$ is approached.  However, the finite discreteness of the Surface Evolver mesh calculations limits the ability to strictly resolve the shape of the hinge below some value of $\Delta \bar{v}$.  Nevertheless, we observe reasonable agreement with the respective power law growth and decay of hinge curvature and size predicted by the asymptotic argument above, particularly for the larger $\Phi$ and intermediate regime of $\Delta \bar{v}$

In Figs.~\ref{fig: nrlyisoSE}C-D we turn to the energetics of fluid-planar inclusion composites in the high inflation regime.  In general, for variable $\Phi$ discrete vesicle simulations agree well with axisymmetric solutions in the regimes where the latter can be resolved.  Fig.~\ref{fig: nrlyisoSE}D tests the prediction for the power-law divergence of bending energy as $\Delta \bar{v} \to 0$.  We note an apparent reciprocal power law regime for each value of $\Phi$ at intermediate values of $\Delta \bar{v}$, but again a true divergence of the bending energy is cut off by the finite energy for crease formation in the triangular mesh approximation to the Helfrich model, eq.~(\ref{eq: SE}).  Also, we find that the slope of dependence in the log-log plot is somewhat below the expected power-law of $\bar{E}_{\rm b} \sim (\Delta \bar{v})^{-1}$, particularly for lower $\Phi$. We note that the slope in this intermediate regime tends somewhat towards $-1$ as $\Phi$ increases, suggesting perhaps that the approximation that $\kappa_{\rm hinge}$ is much larger than the hoop curvature may limit the validity of analysis for insufficiently large $\Phi$ values.

\section{Discussion and Conclusion}\label{sec: conclusion}

We have explored how rigid planar inclusions reshape the ground state symmetries and energetics of vesicles.  In particular, we find that circular inclusions strongly shift the balance of stability towards oblate over prolate shapes, an effect that is driven by the cost of redistributing the ``excess'' Gaussian curvature expelled from the inclusion to the fluid and which therefore increases with $\Phi$.  While we find that this same effect drives ground states to break axisymmetry for low $\Phi$ and sufficiently low inflation shapes, ground states retain axisymmetric, oblate shapes for in the high-inflation regime, and for planar fractions larger than about 10\% of the vesicle area.  Notably, this results in shifting conditions for zero-tension (i.e. minimal bending energy) considerably downward in reduced volume with increasing $\Phi$.  

One direct way to assess the relevance of this result is comparison to the case of fluid-solid composite vesicles possessing a single large and relatively compact 2D solid domain.  In Fig.~\ref{fig: microscopic} we show some example vesicles prepared from mixed DPPC and DOPC lipids prepared via a combination of electroformation and controlled heating and recooling, as is described elsewhere\cite{bandekar2012floret,chen2014large} (see Appendix~\ref{app: exp} for details specific to this results).  Fig.~\ref{fig: microscopic}A-C show three different example vesicles, extracted from videos allowing for perspectives on different orientations and focal sections.  Notably, the fluorescent dye is present in the fluid domain, and the single 2D solid domain appears dark.  Comparison to oblate shape equilibria for axisymmetric fluid-planar inclusions shows apparently good agreement for the 3D shapes and provides a means to estimate the relevant values of $\Phi$ and $\bar{v}$.  It is notable the values of $\bar{v}$, particularly for Fig.~\ref{fig: microscopic}A-B, are quite low, with the lowest value showing obvious and pronounced inward curvature opposite to the solid domain.  Such highly underinflated, quasi-discoidal, shapes are otherwise difficult to stabilize in homogeneous vesicles, at least without significant inner/outer asymmetry, and the fact that they are long-lived in fluid-solid composites is suggestive of the importance of the significant shifting of the zero-tension state downward in reduced volume for $\Phi \approx 0.1$.

Based on the assumption of planarity for the inclusion, in Sec.~\ref{sec: nearlyiso} we analyzed the build up of tension in oblate shapes for $\bar{v} < \bar{v}_*(\Phi)$ and show that energetics derive from a concentration of bending energy at the hinge whose size becomes decreasingly narrow in the asymptotic limit $\bar{v}\to \bar{v}_{\rm max}$ as the isoperimetric limit volume is approached.  We note that the effect of increasing inflation on the hinge region at the boundary of solid domains has previously been considered in fluid-solid composites possessing multiple solid domains~\cite{xin2021switchable}.  In this case, it was argued that large bending energies at the edges of solids favored at least partial overlaps of hinges, in effect a ``depletion-like'' mechanism that gives rise to inter-domain attraction, with the spacing controlled to the hinge size for sufficiently inflated vesicles.  In ref.~\cite{xin2021switchable} a 2D vesicle model was used to study inter-domain energetics for pair-wise domain interactions.  In this context, we expect that the asymptotic understanding of scaling dependence of hinge structure in nearly isoperimetric vesicles may be valuable for extending models of multidomain interactions to 3D vesicles beyond the two-domain case.

The overall good correspondence between strongly oblate shapes and experimentally observed fluid-solid vesicles and the ground shapes of the planar inclusion model suggests the basic shape frustration by the compact 2D solid domain is well approximated by a strictly rigid plane, at least in the underinflating regime shown in Fig.~\ref{fig: microscopic}.  It is important to note that in this system, the solid domain is not directly visible, and moreover, while it is reasonable to expect solid elasticity to resist Gaussian curvature (at least in the absence of topological defects in the solid) thin 2D solids are still highly flexible without changes in Gaussian curvature, i.e. isometric deformations.  More generally, it can be argued that for sufficiently large values of $R/t$ solid domains should be fairly free to adopt zero Gaussian curvature shapes with non-zero mean curvature, known as {\it developable surfaces}~\cite{Witten_2007}.  This class of shapes was shown to be relevant to fluid-solid vesicles in which the solid possesses highly-elaborate, non-convex flower-like shapes\cite{wan2023flowering}.  Surface Evolver simulations showed that for sufficiently high inflation, when the vesicles are subject to large positive tensions, the outer periphery of solid domains tends to bend cylindrically before conforming to the gross spherical shape of the fluid vesicle.  Notably, the geometric rules for developable folding of non-zero Gaussian curvature shapes are fairly restrictive\cite{starostin2007shape}, requiring straight foldlines (known as generators) that extend through the entire domain and without crossing.  Even for circular domains, therefore, it can be expected that optimal deformations consist of quasi-polygonal fold patterns circumscribed by the boundary of the solid domain itself.  This suggests an entirely distinct, and currently unexplored, mechanism for breaking axisymmetry in ground state shapes of fluid-solid composite vesicles whose optimal patterns will be controlled by both solid area fraction and vesicle inflation.

\section*{Acknowledgements}
We thank B. Davidovitch and P. Ziherl for useful discussions about aspects of this work.  This work was supported by the U.S. Department of Energy, Office of Science, Basic Energy Sciences, under award DE-SC0017870.  Surface Evolver simulations were performed on the Unity Cluster at the Massachusetts Green High Performance Computing Center

%%%END OF MAIN TEXT%%%


%\balance


\appendix

\section{Axisymmetric shape equations}

\label{app: axi}
The axisymmetric expression of the bending energy reads\cite{seifert1991shape}
\begin{equation}
    E_\text{b}=2\pi \int{\rm d} s\ \frac B2 X\left(\psi'+\frac{\sin\psi}{X}\right)^2,
\end{equation}
and the elastic free energy and shape equation become
\begin{multline}
    F=2\pi B\int{\rm d} s\ \big[\frac 12 X\left(\psi'+\frac{\sin\psi}{X}\right)^2+\bar \Sigma X \\ -\frac{\bar P}{2}X^2\sin\psi+\gamma(X'-\cos\psi)\big]
\end{multline}
and
\begin{equation}\label{axishapeeq}
    \begin{split}
        &\psi''=-\frac{\psi'}{X}\cos\psi+\frac{\cos\psi\sin\psi}{X^2}+\frac{\gamma}{X}\sin\psi-\frac{\bar P X}{2}\cos\psi\\
        &\gamma'=\frac{\psi'^2}{2}-\frac{\sin^2\psi}{2X^2}-\bar PX\sin\psi+\bar \Sigma\\
        &X'=\cos\psi
    \end{split}
\end{equation}
respectively, where prime($'$) denotes a derivative with respect to $s$, $\bar P\equiv P/B$ and $ \bar \Sigma\equiv\Sigma/B$ are rescaled Lagrange multipliers, and $\gamma$ is another Lagrange multiplier introduced to impose the geometric relations among $X, Z$, and $\psi$;
\begin{equation}
    \begin{split}
        &X'=\cos\psi\\
        &Z'=\sin\psi
    \end{split}
\end{equation}
The boundary conditions are as follows
\begin{equation}\label{bcs}
    \begin{split}
        &X(0)=0,\qquad \psi(0)=0\\
        &X(s_\text{max})=\frac D2,\qquad \psi(s_\text{max})=\pi
    \end{split}
\end{equation}
where $D$ is the diameter of the circular inclusion. The conditions on $\psi$ are to prevent diverging bending energy along the boundary and at the bottom. At the maximum inflation $\bar v_\text{max}$, the boundary angle doesn't satisfy eq.~(\ref{bcs}), \textit{i.e.,} $\psi(s_\text{max})=\pi-\theta_\text{iso}\ne\pi$ and therefore the bending energy diverges.

% \begin{figure}[t!]
% \begin{center}
% \includegraphics[width=0.7\columnwidth]{isoperimetric.jpg}%
% \end{center}
% \caption{\label{isoperimetric}The isoperimetric limit configuration. $\theta_\text{iso}(\Phi)$ is the tangent angle at the boundary.}
% \end{figure}

We numerically evaluated eq.~(\ref{axishapeeq}) with the boundary conditions eq.~(\ref{bcs}) changing $\Phi$ and $\bar v$ to find the equilibrium shapes and their elastic energies.

\section{Surface Evolver minimization}
\label{app: SE}
The computations and simulations were performed in the computational software SE. The mesh was made with $\approx44,000$ vertices, and the planar inclusion was imposed by spatially freezing the vertex positions. The bending energy was computed by the built-in function ``\textit{star\_perp\_sq\_mean\_curvature}'', and the fluid membrane configuration was optimized by using \textit{gradient descent} and \textit{Hessian method} until it got fully equilibrated for given sets of physical constraints; $\Phi$ and $\bar v$. Some auxiliary commands such as \textit{vertex averaging} and \textit{jiggle} were done as needed\cite{brakke1992surface,brakke2013surface}. The detailed protocol is as follows
\begin{itemize}
    \item[1)] Prepare a spherical mesh with unit volume (in SE units) centered at the origin. One can start from ``cube.fe" in the Surface Evolver program directory~\cite{brakke2013surface} and minimize the area at constant volume, using steepest decent (``g") and ``hessian\_seek", while sequentially refining the mesh (``r") and vertex averaging (``g"), until target mesh resolution is reached and energy is fully relaxed to spherical shape.  Meshes used in these simulations these required 6 iterations of refinement.

    \item[2)] Next, the isoperimetric perimetric intial state is prepared by ``flattening'' a portion of that mesh above a target height $h$ (relative to centroid of the sphere) down to a plane at that constant vertical position.  For a given target value of planar fraction $\Phi_\text{target}$ this corresponds to a target value, $h(\Phi_\text{target})/R_0 = (1 - 3\Phi_\text{target})/(1 - \Phi_\text{target})$, where $R_0$ is the initial spherical radius.  For each vertex coordinate whose vertical position $z_\alpha>h(\Phi_\text{target})$, its vertical position is reset to $z_\alpha=h(\Phi_\text{target})$ using the ``set'' command.  Due to discreteness, the edge of this projected region does not conform to a circle.  To improve the resolution at the boundary, vertices in this planar region close to the circular edge, are displaced to exactly a radial distance $D/2$ from the pole.  This is done by defining a the constraint that $x_\alpha^2+y_\alpha^2 = (D/2)^2$ for those vertices with $z_\alpha = h(\Phi_\text{target})$ and $x_\alpha^2+y_\alpha^2 -(D/2)^2 <0.005 $ (i.e. within a narrow distance from the edge).
    
%    Project the upper vertices of which $z$ coordinates are larger than $h(\Phi_\text{target})$ down to $h(\Phi_\text{target})$;
%    \begin{itemize}
%        \item foreach vertex vv where vv.z>=z0 do set vv.z h($\Phi_\text{target}$)
%    \end{itemize}
    
%    $h(\Phi_\text{target})$ can be calculated as
%    \begin{equation}\begin{split}
%       & h(\Phi_\text{target})\\
%        &=\frac{R_0}{\Phi_\text{target}-1}\left[\Phi_\text{target}-\sqrt{\Phi_\text{target}^2+(3\Phi_\text{target}-1)(\Phi_%\text{target}-1)}\right]        
 %   \end{split}
 %   \end{equation}
 %   where $R_0$ is the radius of the sphere, however, for this discrete setup, you can not avoid some mismatch and one can manually micro-tune $h(\Phi_\text{target})$ for a better fit.
    
    %Likewise, due to the discreteness, the boundary of the top essentially doesn't make a perfect circle. To improve the %resolution, define a thin layer around the boundary circle and push all the vertices within such layer to a perfect circle %defined by the \textit{level-set constraint} as follows;
    %\begin{itemize}
        %\item define CONSTRAINT 1 in the mesh file;
        
        %FUNCTION:  (x)\^{}2 + (y)\^{}2 = (3/(4*Pi))\^{}(2/3)-h($\Phi_\text{target})$\^{}2
        %\item apply CONSTRAINT 1 on top vertices
        
        %foreach vertex vv where vv.z=h($\Phi_\text{target}$) and (vv.x\^{}2 + vv.y\^{}2 + vv.z\^{}2)\^{}(3/2) >= 3/(4*Pi) - 0.005 %do set vv constraint 1    
  %  \end{itemize}
    
    \item[3)] We identify the planar inclusion and the fluid membrane, by setting the ``color" attribute of facets in the planar inclusion (e.g. to ``green''), while remaining facets in the are set to a different color (e.g. to ``white''), in the Surface Evolver mesh file.  In practice, a facet is in the planar region if all three vertices satisfy $z_\alpha = h(\Phi_\text{target})$.  Using this color attribute, quantities are defined to measure the areas of solid and fluid portions from the facets belonging to these color categories.  For the planar faces vertices are set to ``fixed''.

 %   \begin{itemize}
 %       \item set facet ff color green where min(ff.vertex,z)=z0
 %       \item set facet quantity planar\_area where color==green
 %       \item foreach facet ff where color==green do set ff.vertex fixed
 %       \item set facet quantity fluid\_area where color==white
 %   \end{itemize}

    \item[4)] Set the target fluid area and total volume to match $\Phi_\text{target}$ and $\bar v_\text{target}$.
%    \begin{itemize}
%        \item set facet tension 0
%        \item fix fluid\_area
%        \item fluid\_area.target:=($1-\Phi_\text{target}$)*domain\_area.value/ $\Phi_\text{target}$
%        \item fix vol
%        \item vol.target:=(4*Pi/3)*$\bar v_\text{target}$*(4*Pi/(domain\_area. value+fluid\_area.target))\^{}(-3/2)
%        \item every time when adjusting the target volume, repeat 100 ``V'' and 100 ``g'' 3 times
%    \end{itemize}
    
    \item[5)] Relax the fluid bending energy using 100 ``conjugate\_grad'' and ``hessian\_normal'' relaxation steap followed by ``hessian\_seek" until step size falls below threshold ($10^{-9}$), followed by ``saddle'' check operation.  This is repeated three times, followed by relaxation without ``conjugate\_grad''.  

\end{itemize}

\begin{figure}[t!]
\begin{center}
\includegraphics[width=0.65\columnwidth]{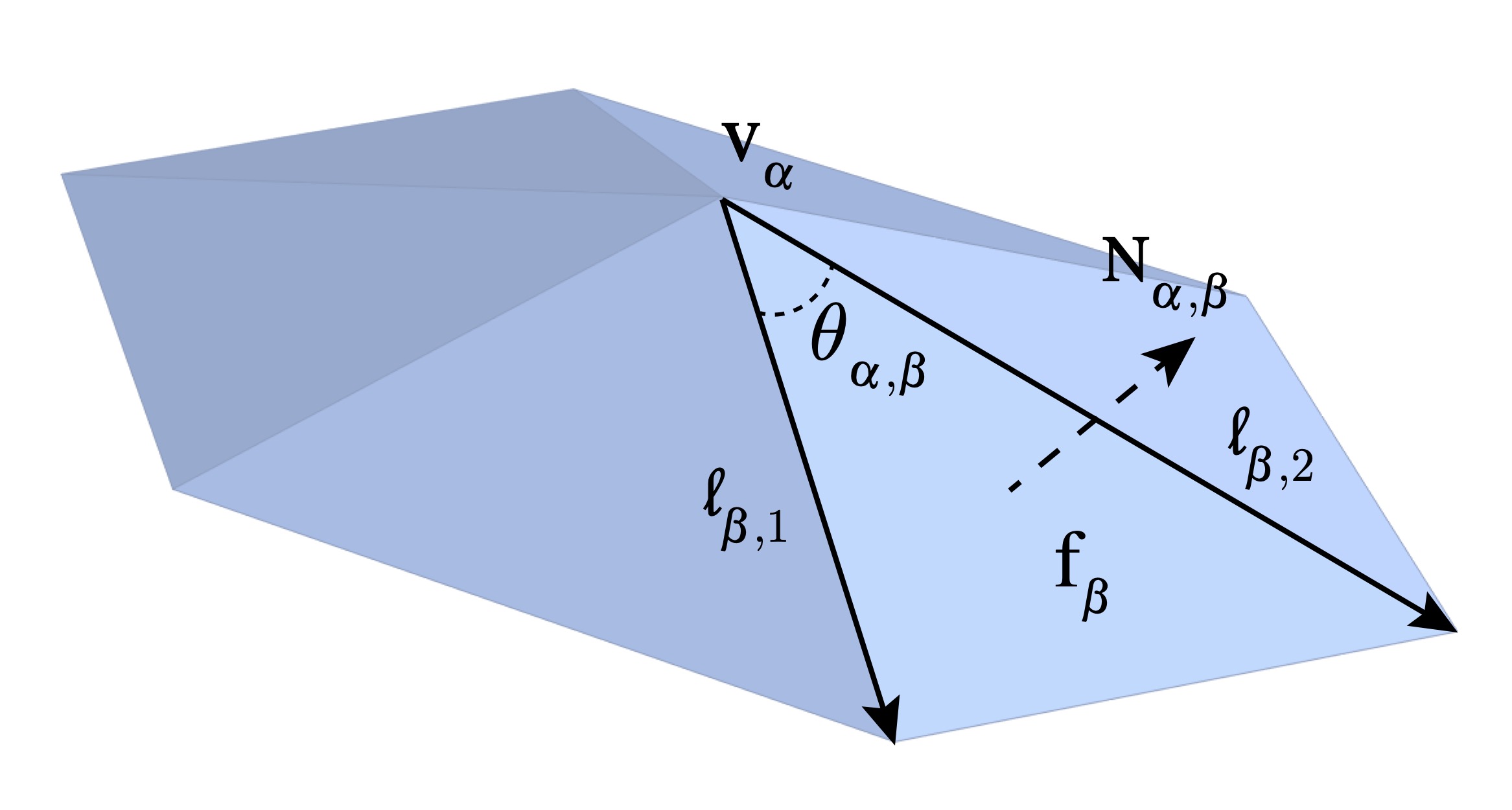}%
\end{center}
\caption{\label{discrete} Schematic of local mesh geometry used in discrete curvature calculations.}
\end{figure}

The bending energy of SE models is given in eq. (\ref{eq: SE}) where
\begin{equation}
H_\alpha=\frac12\frac{\mathbf{F}_\alpha\cdot\mathbf{N}_\alpha}{{\mathbf{N}_\alpha\cdot\mathbf{N}_\alpha}}
\end{equation}
is approximated mean curvature at vertex $\bar v_\alpha$. Here
\begin{equation}\begin{split}
    \mathbf{F}_\alpha&=\sum_\beta\nabla_{\mathbf{v}_\alpha}A_{\alpha,\beta}\\
    &=\sum_\beta\frac{1}{4A_{\alpha,\beta}}\big[-\ell_{\beta,1}|\mathbf{\ell}_{\beta,2}|^2-|\mathbf{\ell}_{\beta,1}|^2\mathbf{\ell}_{\beta,2} \\ & \ \ \ \ \ \ \ \  \ \ \ \ \ \ \ \ \ \ \ \ \ \ \ \  +(\mathbf{\ell}_{\beta,1}\cdot\mathbf{\ell}_{\beta,e})(\mathbf{\ell}_{\beta,1}+\mathbf{\ell}_{\beta,2})\big]
\end{split}
\end{equation}
is the area gradient,
\begin{equation}
    \mathbf{N}_\alpha=\sum_\beta\frac13 A_{\alpha,\beta}\mathbf{N}_{\alpha,\beta}
\end{equation}
is the average normal, and
\begin{equation}
    \Delta A_\alpha=\sum_\beta A_{\alpha,\beta}
\end{equation}
is effective area at vertex $\mathbf{v}_\alpha$, where $A_{\alpha,\beta}$ and $\mathbf{N}_{\alpha,\beta}$ are the area and unit normal at the adjacent facets $\bf{f}_\beta$\cite{meyer2003discrete,crane2018discrete} [See Figure~\ref{discrete}].

\section{Formation and imaging solid-fluid states of DOPC:DPPC vesicles}

\label{app: exp}
Here we summarize the methods used to generate and image fluid-solid composite vesicles, each containing a single, compact 2D solid domain within its otherwise fluid membrane. Methods to produce the giant unilamellar vesicles are previously described in detail~\cite{chen2014large, wan2023flowering}.

DOPC (1,2-dioleoyl-sn-glycero-3-phosphocholine), DPPC (1,2-dipalmitoyl-sn-glycero-3-phosphocholine), and fluorescent tracer lipids Rh-DOPE (1,2-dioleoyl-sn-glycero-3-phosphoethanolamine-N-(lissamine rhodamine B sulfonyl)) were purchased from Avanti Polar Lipids (Alabaster, AL). Vesicles having a 30/70 (Fig.~\ref{fig: exam}A and Fig.~\ref{fig: microscopic}B,C) and 20/80 (Fig.~\ref{fig: microscopic}A) weight ratio DPPC/DOPC plus 0.1 mol\% Rh-DOPE were electroformed on platinum wires. The 10 mM sucrose electroforming solution was preheated and the electroforming temperature was maintained in the range 55–70$^\circ$C to ensure vesicles were formed in the one phase region of the phase diagram, all having the same membrane composition. After electroformation, the stock vesicle suspension was harvested and allowed to cool to room temperature for later use, within 2-3 days. Single solid domains were formed by dilution into a closed chamber made from two coverslips and parafilm spacers, reheating to 55$^\circ$C for 5 minutes, and imposing controlled cooling at 0.3$^\circ$C /min from 45$^\circ$C to room temperature.

Vesicles having solid domains were observed using a Nikon Eclipse TE 300 inverted epifluorescence microscope equipped with a 40$\times$ long working distance air fluorescence objective. Images were recorded with a pco.panda 4.2 sCMOS monochrome camera and analyzed using Nikon NIS Elements imaging software.

%%%REFERENCES%%%
\bibliography{reference} %You need to replace "rsc" on this line with the name of your .bib file
\bibliographystyle{apsrev4-2} %the RSC's .bst file

\end{document}